\documentclass[aps,pra,twocolumn,superscriptaddress]{revtex4-1} %twocolumn

\usepackage{graphicx,subfigure}
\usepackage{amsmath,amssymb}
\usepackage{mathtools}
\usepackage{multirow}
\usepackage{textcomp}
\usepackage{float}
\usepackage{xcolor}
\usepackage[normalem]{ulem}
\usepackage{dsfont}
\usepackage{tabularx}
\usepackage{bbold}
\usepackage{colortbl}
\usepackage{placeins}
\usepackage{subfiles}

\usepackage{etoolbox}

\makeatletter
\newcommand*{\newbibstartnumber}[1]{%
	\apptocmd{\thebibliography}{%
		\global\c@NAT@ctr #1\relax
		\addtocounter{NAT@ctr}{-1}%
	}{}{}%
}
\makeatother

\newcolumntype{Y}{>{\center\arraybackslash}X}

\usepackage{hyperref}
\hypersetup{colorlinks=true, urlcolor = blue}

\newcommand{\avg}[1]{\left\langle #1 \right\rangle}

\newcommand{\ket}[1]{\ensuremath{\left\vert{#1}\right\rangle}}

\newcommand{\abs}[1]{\ensuremath{\left\vert{#1}\right\vert}}

\newcommand{\micro}[1]{\ensuremath{\mu\mathrm{#1}}}

\renewcommand{\micro}[1]{\ensuremath \mu\mathrm{#1}}

\renewcommand{\vec}[1]{\ensuremath{\mathbf{#1}}}

\newcommand{\Gsc}{\ensuremath{\Gamma_\mathrm{sc}}}

\newcommand{\Cov}[2]{\ensuremath{\mathrm{Cov}\left(#1,#2\right)}}
\newcommand{\Corr}[2]{\ensuremath{\mathrm{Corr}\left(#1,#2\right)}}
\newcommand{\Var}[1]{\ensuremath{\mathrm{Var}\left(#1\right)}}

\newcommand{\degr}{\ensuremath{^\circ}}

\newcommand{\Wnull}{\ensuremath{\mathcal{V}}}
\newcommand{\x}{\ensuremath{\mathbf{x}}}
\newcommand{\p}{\ensuremath{\mathbf{p}}}
\newcommand{\nObs}{\ensuremath{\mathcal{N}}}
\newcommand{\nObsP}{\ensuremath{\mathcal{P}}}
\newcommand{\nObsQ}{\ensuremath{\mathcal{Q}}}
\newcommand{\Qoffset}{\ensuremath{D}}

\newcommand{\curlyF}{\ensuremath{\mathcal{F}^x}}
\newcommand{\Gammac}{\ensuremath{\Gamma_\mathrm{coll}}}
\newcommand{\Hpert}{\ensuremath{\mathcal{H}}}
\newcommand{\QFI}{\ensuremath{\mathcal{F}}}

\newcommand{\sq}{\ensuremath{\zeta^2}}
\newcount\colveccount

\newcommand*\colvec[1]{
	\global\colveccount#1
	\begin{pmatrix}
		\colvecnext
		}
		\def\colvecnext#1{
		#1
		\global\advance\colveccount-1
		\ifnum\colveccount>0
		\\
		\expandafter\colvecnext
		\else
	\end{pmatrix}
	\fi
}

\newcommand{\up}{\ensuremath{\uparrow}}
\newcommand{\down}{\ensuremath{\downarrow}}
\newcommand{\phimin}{\ensuremath{\phi_{\mathrm{min}}}}
\newcommand{\sqmin}{\ensuremath{\sq_{\mathrm{min}}}}
\newcommand{\sqmax}{\ensuremath{\sq_{\mathrm{max}}}}
\newcommand{\sqmeas}{\ensuremath{\sq_{\mathrm{meas}}}}

\begin{document}
\renewcommand{\figurename}{\textbf{Fig.}}
\title{Engineering Graph States of Atomic Ensembles by Photon-Mediated Entanglement}

\begin{abstract}

Graph states are versatile resources for quantum computation and quantum-enhanced measurement. Their generation illustrates a high level of control over entanglement.  We report on the generation of continuous-variable graph states of atomic spin ensembles, which form the nodes of the graph.  The edges represent the entanglement structure, which we program by combining global photon-mediated interactions in an optical cavity with local spin rotations. By tuning the entanglement between two subsystems, we either localize correlations within each subsystem or enable Einstein-Podolsky-Rosen steering.  We further engineer a four-mode square graph state, highlighting the flexibility of our approach.  Our method is scalable to larger and more complex graphs, laying groundwork for measurement-based quantum computation and advanced protocols in quantum metrology.
\end{abstract}

\author{Eric S. Cooper}\thanks{Authors contributed equally}
\author{Philipp Kunkel}\thanks{Authors contributed equally}
\author{Avikar Periwal}\thanks{Authors contributed equally}
\affiliation{Department of Physics, Stanford University, Stanford, California 94305, USA}
\affiliation{SLAC National Accelerator Laboratory, Menlo Park, CA 94025}
\author{Monika Schleier-Smith}
\affiliation{Department of Physics, Stanford University, Stanford, California 94305, USA}
\affiliation{SLAC National Accelerator Laboratory, Menlo Park, CA 94025}

\maketitle

Entanglement is a key resource for enabling quantum computation and advancing precision measurements towards fundamental limits. Crucial to these applications is the ability to controllably and scalably generate quantum correlations among many particles. A leading platform for achieving these ends are systems of cold atoms. Here, entangled states of over 20 atoms, such as cluster states with applications in quantum computation, have been generated by bottom-up approaches using local interactions~\cite{bluvstein2022quantum}.  Conversely, global interactions among $10^2$ to $10^5$ atoms have been applied to prepare collective entangled states, including squeezed states~\cite{esteve2008squeezing,hamley2012spinnematic,mao2023quantum,leroux2010orientation,pedrozo2020entanglement,greve2022entanglement} that enable enhanced precision in clocks~\cite{leroux2010orientation,pedrozo2020entanglement,hosten2016measurement,robinson2022direct} and interferometers~\cite{greve2022entanglement,pezze2018quantum}. Such states, featuring symmetric correlations between all atom pairs, have been generated by collisions in Bose-Einstein condensates~\cite{esteve2008squeezing,hamley2012spinnematic} and by photon-mediated interactions in optical cavities~\cite{leroux2010orientation,pedrozo2020entanglement,greve2022entanglement}.

Atoms in cavities offer a particularly versatile platform for scalable generation of entanglement~\cite{leroux2010orientation,pedrozo2020entanglement,greve2022entanglement, hosten2016measurement, haas2014entangled, barontini2015deterministic}, with a single mode of light serving as an interface for correlating the atoms across millimeter-scale distances. In this setting, entanglement between spatial modes of an atomic gas has been achieved by splitting a global squeezed state into distinct subensembles~\cite{malia2022distributed}, building on past work with optically dense ensembles in free space~\cite{julsgaard2001experimental} and with spinor condensates~\cite{peise2015satisfying, fadel2018spatial, kunkel2018spatially, lange2018entanglement, kunkel2022detecting}. Combining such top-down generation of entanglement with advances in local control and detection~\cite{periwal2021programmable,deist2022superresolution,welte2018photon} provides the opportunity to engineer and probe richer spatial structures of entanglement, with applications in multimode quantum sensing~\cite{shettell2020graph}, multiparameter estimation~\cite{proctor2018multiparameter}, and quantum computation~\cite{raussendorf2001one}.

A paradigmatic class of multimode entangled states are graph states~\cite{briegel2001persistent}, universal resources for quantum computation~\cite{raussendorf2001one} with broader applications in quantum metrology~\cite{shettell2020graph} and
in simulations of condensed-matter physics~\cite{han2007scheme}.  These states, also known as cluster states, derive their name from a graph that defines the entanglement structure, with edges representing correlations between nodes that may represent either individual qubits or continuous-variable degrees of freedom.  Discrete-variable graph states have been generated with superconducting qubits~\cite{gong2019genuine}, trapped ions~\cite{lanyon2013measurementbased}, and Rydberg atoms~\cite{bluvstein2022quantum}, while continuous-variable graph states have been prepared in photonic systems~\cite{asavanant2019generation,larsen2021deterministic}. Hitherto unexplored are opportunities for combining the benefits of light and matter to engineer graph states with flexible connectivity and long-lived information storage in atomic states.

Here, we report on the generation of programmable multimode entanglement in an array of four atomic ensembles coupled to an optical cavity. To control the structure of entanglement, we intersperse global interactions with local spin rotations.  These two ingredients provide control over the strength of entanglement between subsystems and thereby enable a general protocol for preparing graph states.   As a minimal instance, we prepare and characterize a two-mode graph state that exhibits Einstein-Podosky-Rosen (EPR) steering, a strong form of entanglement which is a resource for quantum teleportation and which has previously been demonstrated using collisional interactions in Bose-Einstein condensates~\cite{peise2015satisfying,fadel2018spatial,kunkel2018spatially,kunkel2022detecting}. To illustrate the versatility of our protocol, we further construct a four-mode square graph state. Our work offers a blueprint for scalable generation of resource states for continuous-variable quantum computation and multimode quantum metrology.

\begin{figure*}
\centering
\includegraphics[width=\textwidth]{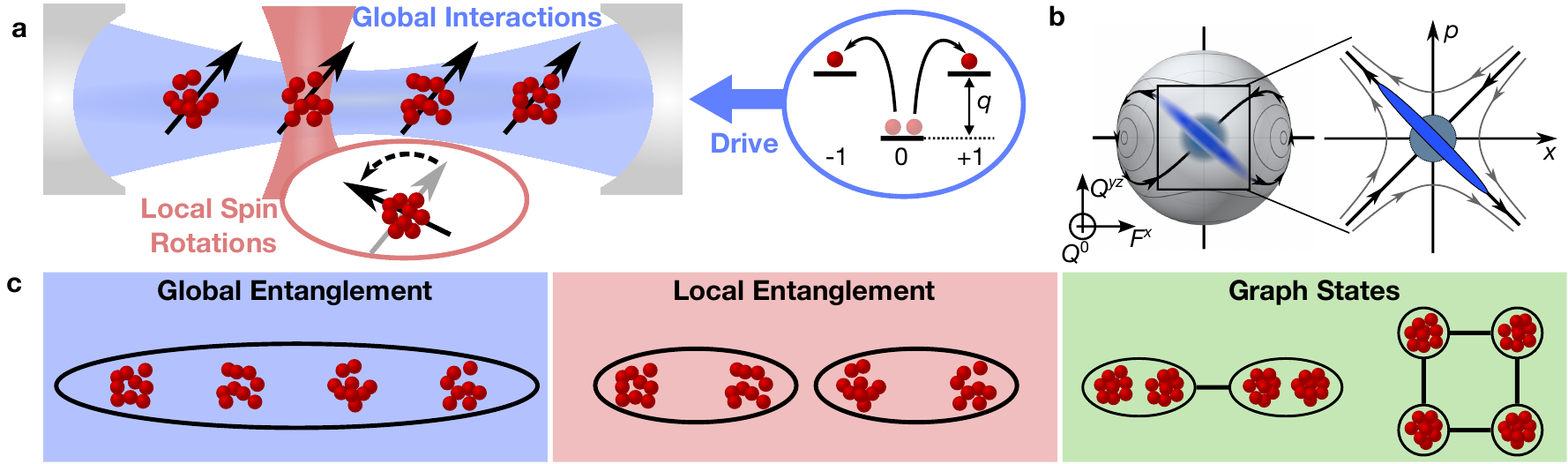}
\caption{\textbf{Programmable entanglement in an array of four atomic ensembles within an optical cavity.}  \textbf{a}, Initializing all atoms in the $m=0$ state and driving the cavity with light induces creation of correlated atom pairs in states $m=\pm 1$.  \textbf{b}, The resulting spin-nematic squeezing is visualized on a spherical phase space spanned by the collective spin-1 observables $\{F^x,Q^{yz},Q^{0}\}$. For short interaction times, the dynamics can be described on an effective two-dimensional phase-space spanned by the conjugate observables $\{x, p\}$.  \textbf{c}, Combining the global interactions with local spin rotations allows for engineering a variety of entanglement structures, such as entanglement localized to selected subsystems and graph states with up to four nodes.
}
\label{fig:overview}
\end{figure*}

As the mechanism for generating global entanglement, we implement cavity-mediated spin-nematic squeezing of spin-1 atoms~\cite{masson2017cavity}. When a drive field is applied to the cavity (Fig.~\ref{fig:overview}a), photons mediate spin-exchange interactions~\cite{davis2019photon}, and the system is governed by the Hamiltonian
\begin{equation}
    H/\hbar = \frac{\chi}{2N} \left(F^xF^x + F^yF^y\right)+ \frac{q}{2} Q^{0}.
    \label{eq:Hamiltonian}
\end{equation}
Here, $\vec{F}$ denotes the collective spin of all $N$ atoms in the cavity, with spin length $F \leq N$, and $\chi$ quantifies the collective interaction strength.  In the second term, $q$ parameterizes the quadratic Zeeman energy, proportional to the difference $Q^0 = N_1+N_{-1}-N_0$ between the populations $N_m$ of atoms in the $m=\pm1$ and $m=0$ Zeeman states.

We visualize the collective spin dynamics in a spherical phase space, analogous to the Bloch sphere, for spin-1 observables (Fig.~\ref{fig:overview}b).  We focus on a system initialized with all atoms in $m=0$, i.e., polarized along the $Q^{0}$ axis.  The effect of the cavity-mediated interactions is to twist the quasiprobability distribution of this initial state about the $F^{x}$ axis, inducing squeezing~\cite{kitagawa1993squeezed}.  Simultaneously, the quadratic Zeeman effect generates so-called spinor rotations about the $Q^{0}$ axis, mapping states along $F^x$ to polarized states of the quadrupole operator $Q^{yz}$ after a rotation of $90\degr$. The early-time dynamics explored in our experiments are well described by approximating a patch of the sphere as a two-dimensional phase space spanned by the conjugate observables $x=F^x/\sqrt{CN}$ and $p=Q^{yz}/\sqrt{CN}$, which are normalized such that the Heisenberg uncertainty relation for $x$ and $p$ is $\Var{x}\Var{p}\geq1$.  The contrast $C$, set by the commutator $\abs{\avg{[F^x, Q^{yz}]}} = 2CN$, accounts for imperfect polarization along the $Q_0$ axis.

We engineer entanglement in an array of four atomic ensembles (Fig.~\ref{fig:overview}a), each containing up to $5\times10^3$ Rubidium-87 atoms in the $f=1$ hyperfine manifold. The ensembles are placed near the center of a near-concentric optical cavity with a Rayleigh range of $0.9~\text{mm}$ and are spaced by $250~\mu\text{m}$. Applying a drive field to the cavity for $50~\mu\text{s}$ generates spin-nematic squeezing in the symmetric mode that directly couples to the cavity.  To read out each ensemble $i$ in a specified quadrature $x_i\cos\phi - p_i\sin\phi$, we map this quadrature onto the spin component $F^x$ via a spinor rotation by an angle $\phi$.  A subsequent spin rotation converts this signal into a population difference between Zeeman states, which we detect by fluorescence imaging.

\begin{figure}[!t]
\centering
\includegraphics[width=\columnwidth]
{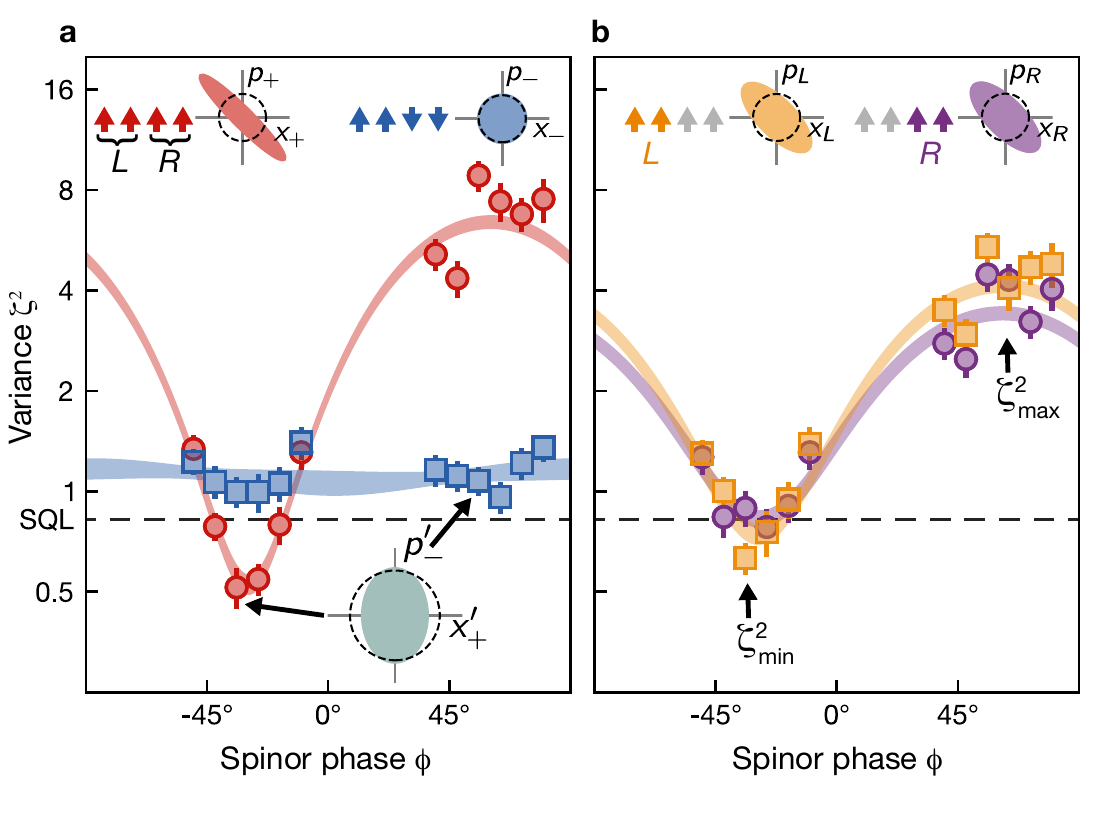}
\caption{\textbf{Global squeezing and entanglement between subsystems.} \textbf{a}, 
Cavity-mediated interactions lead to squeezing of the symmetric mode (red circles) below the standard quantum limit (SQL, dashed line).
The anti-symmetric mode (blue squares) does not couple to the cavity and remains approximately in a coherent state. Multiplying values of the variance $\zeta^2$ for the squeezed quadrature $x_+'$ of the symmetric mode and the orthogonal quadrature $p_-'$ of the antisymmetric mode yields the entanglement witness $W$.  Inset: green ellipse shows area $\sqrt{W}$, smaller than dashed circular region representing minimum-uncertainty unentangled state.
\textbf{b}, Analyzing the left and right subsystems separately (yellow squares and purple circles) yields a degradation in squeezing, consistent with neglecting information contained in correlations between the subsystems.  Error bars show 1 s.d. confidence intervals extracted via jackknife resampling.  Shaded curves show the 1 s.d. confidence intervals of sinusoidal fits to the data.
}
\label{fig:2_global}
\end{figure}

To verify the generation of spin-nematic squeezing, we measure the variance $\sq = \Var{x\cos\phi -p\sin\phi}$ for the symmetric mode $x_+ = \sum_i x_i/2$ of all four ensembles. As shown in Fig.~\ref{fig:2_global}a, we measure a minimum value $\sq = 0.52\pm0.07$, limited primarily by technical noise~\cite{SM}. We confirm the presence of entanglement by evaluating the Wineland squeezing parameter $\xi^2 = \sq/C = 0.63\pm 0.08$. Values below the standard quantum limit (SQL) $\xi^2 = 1$, shown by the dashed line at $\zeta^2 = C$, indicate enhanced metrological sensitivity compared to any unentangled state of $N$ atoms~\cite{pezze2018quantum, SM, wineland1994squeezed}. We calibrate $N$ from measurements of the atomic projection noise (Extended~Data~Fig.~\ref{supp_fig:ShotNoise}) and determine $C$ from measured populations in the three Zeeman states~(Methods Sec.~\ref{sec:contrast}).

To demonstrate that only the symmetric mode couples to the cavity, we also evaluate the variance $\sq$ for the mode $x_- = (x_L-x_R)/\sqrt{2}$ which is anti-symmetric under the exchange of the left two ensembles $x_L$ and the right two ensembles $x_R$. As expected, the variance for the anti-symmetric mode shows no statistically significant dependence on $\phi$ and has an average value $\sq = 1.14\pm0.04$ near the quantum projection noise level.

We confirm the long-range character of the entanglement by evaluating a witness for entanglement~\cite{mancini2002entangling} between the left and right subsystems,
\begin{equation}
    W = \Var{x'_+}\Var{p'_-}.
    \label{eq:entanglement_witness}
\end{equation}
Here, $x'_+$ denotes the squeezed quadrature in the symmetric mode and $p'_-$ is the corresponding conjugate observable in the anti-symmetric mode.
Generically $W$ can take on any value since $x'_+$ and $p'_-$ commute.  However, in the absence of correlations between the left and right subsystems, their independent Heisenberg uncertainty relations impose the constraint $W \ge 1$, such that values $W < 1$ imply entanglement.  The uncertainty product from the data in Fig.~\ref{fig:2_global}a is $W=0.55\pm0.10$, witnessing entanglement between the left and right subsystems. 

Consistent with the entanglement between subsystems, we observe a degradation in squeezing when measuring each subsystem individually, as shown in Fig.~\ref{fig:2_global}b.  To further highlight that the left and right subsystems are in locally mixed states, we quantify the increase in phase space area due to the mutual information between them. For Gaussian states, the phase space area $A_m=\zeta_\text{min} \zeta_\text{max}$ for a mode $m$ is the product of the standard deviations of the squeezed and anti-squeezed quadratures. Local measurements that discard correlations between the left and right subsystems yield a total phase space volume $A_L A_R = 3.7\pm0.4$, larger than the total phase space volume $A_+ A_- = 2.2\pm0.3$ for global measurements of the symmetric and anti-symmetric modes. This emphasizes the loss of information when ignoring correlations between the local subsystems.

\begin{figure*}[!htbp]
\centering
\includegraphics[width=\textwidth]{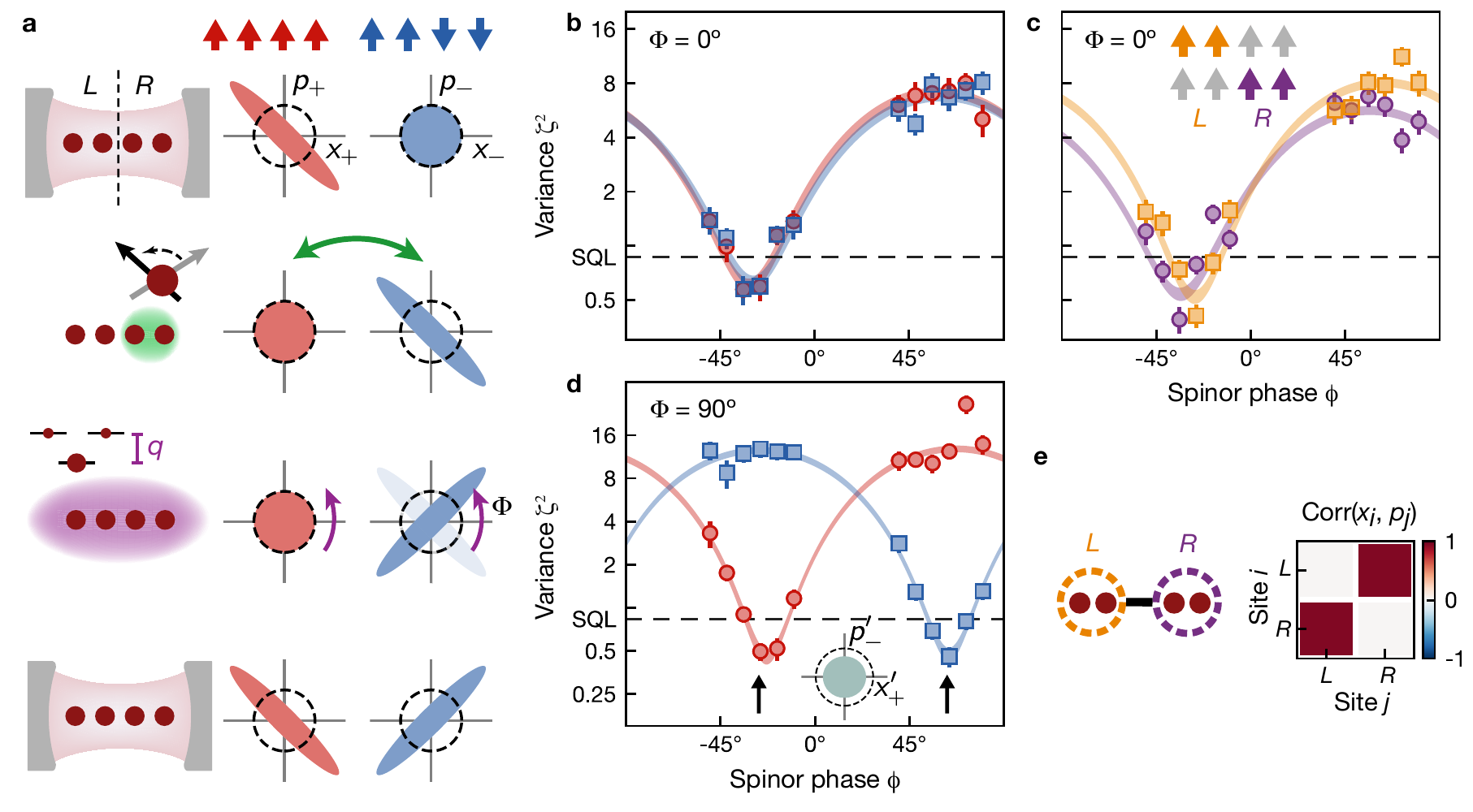}
\caption{\textbf{Tunable entanglement: from local squeezing to EPR correlations.} \textbf{a}, Scheme for controlling the strength of entanglement between left (L) and right (R) subsystems of the four-site array.  After squeezing the symmetric mode (red), we transfer the squeezing into the anti-symmetric mode (blue) by applying a local $180\degr$ spin rotation (green) to the right subsystem. Next, a global spinor rotation (purple) adjusts the angle of the squeezed quadrature. Finally, a second interaction pulse produces squeezing in the symmetric mode. The relative angle $\Phi$  between the squeezed axes of the collective modes determines the form of entanglement. \textbf{b}, To disentangle the left and right subsystems, we choose a relative phase $\Phi=0$ between the squeezed axes of the symmetric (red circles) and anti-symmetric (blue squares) modes.  \textbf{c}, Entanglement internal to each subsystem manifests in variances $\sq = 0.41\pm 0.06$ and $\sq = 0.38\pm 0.07$ for the left and right subsystems (yellow squares and purple circles), respectively. \textbf{d}, To generate EPR entanglement between the left and right subsystems, we choose a relative angle $\Phi =90\degr$ between squeezed quadratures of the collective modes. The variances $\sq = 0.50\pm0.07$ and $\sq = 0.46\pm0.08$ for orthogonal quadratures of the symmetric and anti-symmetric modes yield an entanglement witness $W = 0.23\pm0.05<1$.  \textbf{e}, Representation of the resulting EPR entangled state as a graph state, corroborated by the reconstructed correlation matrix $\Corr{x_i}{p_j}$.}
\label{fig:two_cluster}
\end{figure*}

To optimize squeezing within each subsystem, e.g., for applications in spatially resolved sensing, the correlations between subsystems should be removed while maintaining the entanglement internal to each subsystem. Combining the global spin-nematic squeezing with local rotations provides the requisite control of the entanglement structure. To disentangle the left and right subsystems, we perform a sequence akin to spin echo, as shown in Fig.~\ref{fig:two_cluster}a. Between two pulses of interactions, we rotate the spins of the right subsystem by $180\degr$ by optically imprinting a local vector ac Stark shift (Methods Sec.~\ref{sec:supp_spin_rots}). The effect is to cancel out interactions between the two subsystems, leaving only local squeezing (Fig.~\ref{fig:two_cluster}c).  The scheme can equivalently be viewed as squeezing both the symmetric and anti-symmetric modes in the same quadrature (Fig.~\ref{fig:two_cluster}b). 

More broadly, applying a sequence of squeezing operations in the basis of collective modes enables control over the spatial structure of entanglement via the relative orientations of the squeezed quadratures.  Whereas a relative phase $\Phi = 0$ between the squeezed quadratures of the symmetric and antisymmetric modes disentangles the left and right subsystems, the entanglement between subsystems can alternatively be maximized by introducing a relative phase $\Phi=90\degr$ via a spinor rotation in the sequence shown in Fig.~\ref{fig:two_cluster}a.  The $90\degr$ phase improves the entanglement witness $W$ in Eq.~\eqref{eq:entanglement_witness} by producing simultaneous squeezing of both $x'_+$ and $p'_-$.  The resulting variances, shown in Fig.~\ref{fig:two_cluster}d, yield an entanglement witness $W=0.23\pm0.05$.  The presence of squeezing in both orthogonal quadratures is indicative of entanglement of the paradigmatic Einstein-Podolsky-Rosen (EPR) type.

A notable feature of the EPR entangled state is its capacity for steering, in which measurements of one subsystem can predict measurements of both quadratures of the other subsystem to better than the local Heisenberg uncertainty product.  Steering is a stricter condition than entanglement and enables teleportation of quantum information~\cite{reid2009colloquium}. To witness the left subsystem steering the right, we use measurements of the left subsystem to estimate $x'_R$ and $p'_R$ and calculate the error of the inference after subtracting a small detection noise contribution (see Methods Sec.~\ref{sec:SteeringCriterion}). The product of conditional variances $\Var{x'_R|x'_L}\Var{p'_R|p'_L} = 0.68\pm0.18$ is less than one, the local Heisenberg uncertainty bound.  The comparable witness for the right subsystem steering the left is $0.66\pm 0.18$.  We thus establish bidirectional steering at the $92\%$ confidence level, which justifies identifying the state as a continuous-variable EPR state. 

Our preparation of the EPR state constitutes a minimal instance of a scalable protocol for preparing graph states, in which the edges of the graph denote quantum correlations between conjugate observables on connected sites.  Mathematically, this defining property of an ideal graph state can be expressed as
\begin{equation}
\label{eq:GraphState}
    \Var{p_i - A_{ij}x_j} \rightarrow 0,
\end{equation} 
where the adjacency matrix $A$ encodes the connectivity of the graph and we implicitly sum over the repeated index $j$. As a general recipe for preparing a specified graph state, we diagonalize the adjacency matrix $A$ to obtain a set of eigenvectors representing collective modes that should be squeezed.  For each eigenmode $m$, the corresponding eigenvalue $\lambda_m$ specifies the orientation $\phi_m = \mathrm{arccot}\,\lambda_m$ of the squeezed quadrature.

The graph representing the two-mode EPR state is shown in Fig.~\ref{fig:two_cluster}e and corresponds to an adjacency matrix 
\begin{equation}\label{eq:EPR_graph}
A = \begin{bmatrix}
0 & 1 \\
1 & 0
\end{bmatrix}.
\end{equation}
Diagonalizing $A$ yields a state preparation protocol that matches the scheme of Fig.~\ref{fig:two_cluster}a: the eigenmodes of $A$ are the symmetric and antisymmetric modes, while the eigenvalues $\lambda_\pm = \pm 1$ indicate that the squeezed quadratures should be oriented at $\phi_\pm = \pm 45 \degr$, consistent up to a global rotation with the squeezing curves in Fig.~\ref{fig:two_cluster}d.  Henceforth we work in a globally rotated basis chosen to orient the squeezed quadratures at the angles $\phi_m$.  To visualize the equivalence of squeezing the collective modes with engineering the graph of entanglement, we use the data from Fig.~\ref{fig:two_cluster}d to reconstruct the correlations between conjugate variables in the two subsystems
\begin{equation}
    \label{eq:Corr_xp}
    \Corr{x_i}{p_j} = \frac{\Cov{x_i}{p_j}}{\sqrt{\Var{x_i}\Var{p_j}}},
\end{equation}
where $\Cov{\cdot}{\cdot}$ denotes the covariance (Methods Sec.~\ref{sec:CorrelationMatrix}). These correlations, shown in Fig.~\ref{fig:two_cluster}e, agree with the adjacency matrix $A$. 

\begin{figure*}[ht!]
\centering
\includegraphics[width=\textwidth]{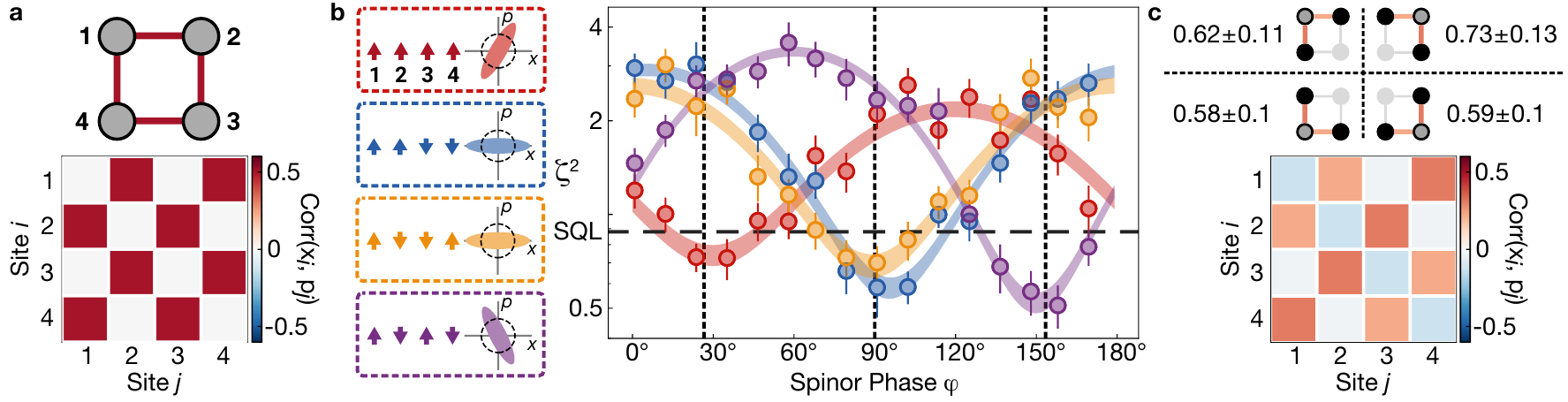}
\caption{\textbf{Generation of a square graph state.} \textbf{a}, Diagram of four-mode square graph state and theoretical correlation matrix $\Corr{x_i}{p_j}\propto A_{ij}$. 
\textbf{b}, Left: schematic illustration of eigenmodes of the adjacency matrix $A$ and the corresponding squeezing ellipses, with orientations specified by eigenvalues $\lambda_m = \cot\,\phi_m$.  Right: measured variances $\zeta^2$ in the four eigenmodes, showing squeezing at the specified spinor phases $\phi_m$ (black dashed lines).  Error bars show 1 s.d. confidence interval.
\textbf{c}, Top: directly measured variances $v_i$ of the nullifiers, with schematics showing central node $i$ (dark gray circle) and neighbors (black circles) contributing to each nullifier. Bottom: correlation matrix reconstructed from the measurement results in \textbf{b}.}
\label{fig:cluster_state}
\end{figure*}

We additionally directly probe the graph of the EPR state by measuring the variances of the nullifers $n_i = p_i - A_{ij}x_j$ in Eq.~\eqref{eq:GraphState}.  As the ideal limit of zero variance requires infinitely strong squeezing, a practical definition of a graph state is that the variances of the nullifiers should approach zero in the limit of perfect squeezing. %as $\zeta \rightarrow 0$ for some variance $\zeta^2$.   
Defining normalized variances 
\begin{equation}
v_i = \Var{n_i}/(1 + \sum_j A_{ij}^2)
\end{equation}
such that $v_i = 1$ for a coherent state, our state preparation protocol theoretically produces variances $v_i = \zeta^2$ assuming equal squeezing of all eigenmodes.  Experimentally, we access each nullifier $n_i$ by performing a local $90\degr$ spinor rotation on subsystem $i$.  For the two-mode EPR state, with $n_L = p_L - x_R$ and $n_R = p_R - x_L$, we measure variances $v_L = 0.53\pm 0.11$ and $v_R = 0.36\pm 0.09$ (Extended Data Fig.~\ref{fig:supp_nullifier_epr}),  directly confirming the entanglement structure specified by the graph.

To illustrate the scaling to more complex graphs, we produce the square graph state shown in Fig.~\ref{fig:cluster_state}a, with adjacency matrix
\begin{equation}
    A = \begin{bmatrix}
    0 & 1 & 0 & 1\\
    1 & 0 & 1 & 0\\
    0 & 1 & 0 & 1\\
    1 & 0 & 1 & 0
    \end{bmatrix}.
\end{equation}
The eigenbasis of $A$ is shown in Fig.~\ref{fig:cluster_state}b.  The eigenvalues $\lambda_m = \left(2, 0, 0, -2\right)$ specify squeezing angles $\phi_m = \left(27\degr, 90\degr, 90\degr, 153\degr\right)$ for the four eigenmodes. We sequentially couple each eigenmode to the cavity with the aid of local spin rotations, analogously to the scheme in Fig.~\ref{fig:two_cluster}a, squeezing the desired quadrature of each mode via global cavity-mediated interactions followed by a global spinor rotation (Extended Data Fig.~\ref{fig:supp_sequence}).  The result is shown in Fig.~\ref{fig:cluster_state}b, where the orientation of the squeezed quadrature for each eigenmode is within $5\degr$ of the target squeezing angle $\phi_m$.  Reconstructing the correlations $\Corr{x_i}{p_j}$ between sites from these measurements of the collective modes yields the matrix shown in Fig.~\ref{fig:cluster_state}c, which is consistent with the target adjacency matrix.

We additionally directly measure the nullifiers $n_i$ for the square graph state. Their normalized variances $v_i$, listed in Fig.~\ref{fig:cluster_state}e,  have an average value $0.63\pm0.07$ consistent with the squeezing $\zeta^2$ of the collective modes.   Each nullifier further satisfies a condition $v_i < 0.94$ ruling out separability into the independent nodes of the graph (Methods Sec.~\ref{sec:entanglement_detection}), highlighting the presence of spatial entanglement between the four ensembles.

Our scheme for preparing graph states generalizes to any method of generating global entanglement that can be combined with local rotations.   The approach is scalable to larger arrays, requiring only $M$ squeezing operations to prepare arbitrary $M$-node graph states.  For atoms in a cavity, the rate of each squeezing operation is collectively enhanced by the number of modes, such that the total interaction time required is independent of array size~\cite{SM}.  Similarly, the degree of squeezing per mode is fundamentally limited only by the collective cooperativity per ensemble.  Combining our approach with cavity-mediated generation of non-Gaussian states~\cite{haas2014entangled, barontini2015deterministic,colombo2022time} or atom counting~\cite{bourassa2021blueprint, hume2013accurate, bochmann2010lossless, gehr2010cavity} opens prospects in continuous-variable quantum computation.  Proposals for fault-tolerant measurement based quantum computation with continuous-variable cluster states require initial squeezing of 20.5~dB~\cite{menicucci2014fault}, which is comparable to the strongest demonstrated cavity-based spin squeezing~\cite{hosten2016measurement}. Continuous-variable cluster states additionally enable novel forms of quantum-enhanced measurement~\cite{shettell2020graph, proctor2018multiparameter}, including simultaneous sensing of displacements in conjugate variables~\cite{tsang2012evading} with applications including vector magnetometry. 

Our protocol can be extended to a variety of platforms where either bosonic modes or qubits form the nodes of the graph and a central ancilla mediates collective interactions.  Opportunities include generating continuous-variable graph states in multimode optomechanical systems~\cite{piotrowski2023simultaneous}, or in superconducting circuits featuring multiple microwave or acoustic modes coupled to a single qubit~\cite{hann2019hardware,naik2017random}; and discrete-variable graph states of individual atoms, superconducting qubits~\cite{zhang2023superconducting}, or ions~\cite{bohnet2016quantum} with photon- or phonon-mediated interactions.   Our approach offers the benefit of programmable connectivity and prospects for leveraging the central ancilla to perform quantum non-demolition measurements with applications in computation, error correction, and continuous quantum sensing.

\section*{Methods}

\setcounter{table}{0}
\renewcommand{\tablename}{\textbf{Extended Data Table}}
\renewcommand{\thetable}{\textbf{\arabic{table}}}

\setcounter{figure}{0}
\renewcommand{\figurename}{\textbf{Extended Data Fig.}}
\renewcommand{\thefigure}{\textbf{\arabic{figure}}}
\renewcommand{\theHfigure}{extdata.\arabic{figure}}
\renewcommand{\theHtable}{extdata.\arabic{table}}

\subsection{Definition of spin and quadrupole operators}
\label{sec:definition}
While for spin-1/2 particles all single-particle spin operators can be written as a linear combination of the dipole moments $f^x$, $f^y$, and $f^z$, the space of spin-1 operators additionally includes quadrupole operators defined as $q^{\alpha\beta} = f^\alpha f^\beta + f^\beta f^\alpha - \frac{4}{3}\delta_{\alpha\beta}$ where $\alpha, \beta \in \{x,y,z\}$ and $\delta_{\alpha\beta}$ is the Kronecker delta function~\cite{hamley2012spinnematic}. For plotting the state on the generalized Bloch sphere, we use the operator $q^0 = q^{zz} + \frac{1}{3}$, which quantifies the population difference between the $m=0$ state and the $m=\pm1$ states.  We additionally construct collective observables $F^{\alpha} = \sum_i^N f_i^{\alpha}$ and ${Q^{\alpha\beta} = \sum_i^N q_i^{\alpha\beta}}$ corresponding to each spin-1 operator in a system of $N$ atoms. 

\subsection{State Preparation}
\label{sec:experiment}
To prepare the array of four atomic ensembles in an optical cavity, we initially load $^{87}\mathrm{Rb}$ atoms from a magneto-optical trap into an array of optical dipole traps, each with a waist of 6$~\mu$m. After optically pumping the atoms into the $\ket{f = 2, m = -2}$ state, the ensembles are transferred into a 1560 nm intracavity optical lattice. Further details of the trapping procedue are described in Ref.~\cite{periwal2021programmable}.  The atoms are then evaporatively cooled by decreasing the lattice depth from $U_0=h\times 14$~MHz to $U_0=h\times 175$~kHz in 200~ms. A series of composite microwave pulses~\cite{levitt1986composite} is used to transfer the atoms from $\ket{2, -2}$ to $\ket{1, 0}$.  Any remaining population in the $\ket{1, \pm 1}$ states is removed by first transferring this population into the $\ket{f = 2}$ manifold using microwave pulses, and then applying resonant light to push and heat the $\ket{f = 2}$ population out of the lattice.  The lattice is then ramped up to a depth of $U_0 = h\times 25$~MHz to minimize atom loss and increase confinement during the interaction phase of the sequence, yielding a final temperature in the lattice of 80~$\mu$K.  During the interaction phase of the experiment, the ratio of the lattice depth to atomic temperature is $U_0/(k_B T) = 15$ for an ensemble at the center of the cavity.

\subsection{Interactions and Cavity Parameters}
\label{supp_sec:interactions}
The spin-exchange interactions between atoms are mediated by a near-concentric Fabry-Perot cavity with length $2R-d$, where $R=2.5~$cm is the radius of curvature of the mirrors and $d =$ 70~$\mu$m. The drive field is detuned from the $\ket{5S_{1/2}, f = 1}\rightarrow\ket{5P_{3/2}}$ transition by $\Delta = -2\pi\times 9.5$~GHz, after accounting for the ac Stark shift of the excited state due to the 1560~nm lattice.  At the drive wavelength of 780~nm, the cavity mode has a Rayleigh range $z_R = 0.93$~mm and waist $w_0=15~\micro{m}$, resulting in a vacuum Rabi frequency $2g = 2\pi\times 3.0$~MHz.  Comparing with the cavity linewidth $\kappa=2\pi\times250$~kHz and atomic excited-state linewidth $\Gamma = 2\pi\times 6.1$~MHz yields a single-atom cooperativity $\eta_0 = \frac{4g^2}{\kappa\Gamma} = 6.1$ for a maximally coupled atom at cavity center.

We parameterize the dispersive atom-light coupling by the vector ac Stark shift per intracavity photon, which for a maximally coupled atom is $\Omega_0 = -\frac{g^2}{6\Delta} = 2\pi\times 41$~Hz.  As the array of atomic ensembles spans a length of 750~$\mu$m along the cavity axis, centered at the focus of the cavity mode, the maximally coupled ensembles experience a 30\% larger Stark shift than the two minimally coupled ensembles.  In addition, thermal motion of the atoms in the lattice means that the average atom experiences a reduced single-photon Stark shift compared with an on-axis atom at an antinode, resulting in a thermally averaged single-photon Stark shift $\Omega = 2\pi\times 27$~Hz.

Our method of generating cavity-mediated interactions is described in Refs.~\cite{davis2019photon,davis2020protecting}.
The interactions are controlled by a drive field detuned from cavity resonance by an amount $\delta_c$.  This corresponds to detunings $\delta_{\pm}=\delta_c\mp\omega_z$ from two virtual Raman processes in which a collective spin flip is accompanied by emission of a photon into a cavity, where $\omega_z$ is the Zeeman splitting.  Rescattering of this photon into the drive mode is accompanied by a second collective spin flip, producing resonant spin-exchange processes of collective interaction strength 
\begin{equation}
    \label{supp_eq:chi_def}
    \chi^{\pm} = N\overline{n}\frac{\Omega^2}{2}\frac{\delta_{\pm}}{\delta_{\pm}^2 + \left(\frac{\kappa}{2}\right)^2},
\end{equation}
where $N$ is the total number of atoms and $\overline{n}$ is the intracavity photon number~\cite{davis2020protecting}. We operate in a magnetic field of $4.1$ G perpendicular to the cavity axis, corresponding to a Larmor frequency $\omega_z = 2\pi\times 2.9$~MHz.  The drive light is typically detuned by $2\pi\times4.2$~MHz from the shifted cavity resonance, so that $\delta_- = -2\pi\times 1.3$~ MHz and $\delta_+ = -2\pi\times 7.1$~MHz.  We define a total interaction strength $\chi = \chi^- + \chi^+$.  The drive light produces a typical intracavity photon number $\overline{n} = 800$.  A representative atom number $N = 1.5\times10^4$ yields a collective interaction strength $\chi = -2\pi\times 4$~kHz. Exact parameters for each data set are detailed in Extended Data Table~\ref{tab:figure_parameters}. The parameters were selected to optimize squeezing, as discussed in Sec.~\ref{sec:squeezing_dynamics} of the supplement.

\subsection{Global and Local Control over Spin Orientation}
\label{sec:supp_spin_rots} 
To access different quadratures of the squeezed states generated in our experiments and to adjust the relative squeezing angles of the collective modes, we apply global rotations about the $Q^{0}$ axis by two different methods. In the first method we let the system evolve under the quadratic Zeeman shift $q=2\pi\times 1.2~$kHz.  Alternatively, we apply a detuned $2\pi$ microwave pulse on the hyperfine clock transtion $\ket{f = 1, m = 0} \leftrightarrow \ket{f = 2, m = 0}$.  For a suitable choice of detuning $\delta_\mathrm{mw}$ and microwave Rabi frequency $\Omega_\mathrm{mw}$, the imparted phase is $\phi = \pi(1-\delta_\mathrm{mw}/\sqrt{\Omega_\mathrm{mw}^2+\delta_\mathrm{mw}^2})$. This latter technique reduces the time required to rotate the orientation of the squeezed state before the final readout, since the Rabi frequency $\Omega_\mathrm{mw}=2\pi\times7.5~$kHz is much larger than the quadratic Zeeman shift. However, inhomogeneities in the microwave Rabi frequency on different ensembles can lead to unwanted population transfer from $\ket{1,0}$ to $\ket{2, 0}$, which shifts the cavity resonance for subsequent interaction periods. Therefore, in sequences employing multiple drive field pulses to squeeze different collective modes, we use only the rotation under quadratic Zeeman shift to adjust the squeezing angle.  

We apply local spin rotations around $F^y$  to read out the observables $x$ and $p$, and rotations about $F^z$ to transform between collective modes. For these rotations, we use circularly polarized light that is blue-detuned from the $\ket{5S_{1/2}, f = 1}\rightarrow\ket{5P_{3/2}}$ transition by 120~GHz. The laser beam is perpendicular to the cavity axes and is focused to individually address a single atomic ensemble, which we select by controlling the position of the beam via an acousto-optical deflector (AOD). The angle between the magnetic field, which defines our quantization axis, and the propagation direction of the laser is chosen to be $70\degr$. The circular component parallel to the magnetic field induces a vector ac Stark shift that acts as an artificial magnetic field, generating local rotations about $F^z$.  Rotations by $180\degr$ about $F^z$ flip the sign of both $F^x$ and $Q^{yz}$ on selected ensembles. We thus utilize these rotations to transfer squeezing between orthogonal collective modes, as shown in Fig.~\ref{fig:two_cluster}a of the main text. For this transfer we simultaneously address two ensembles and induce the required spin rotation in approximately 18~$\mu$s. 

The same laser allows for driving Raman transitions within the $f=1$ hyperfine manifold, as the circular polarization component orthogonal to the magnetic field acts as an effective transverse field.  Specifically, we use an arbitrary waveform generator to modulate the drive amplitude of an acousto-optical modulator, and thus the power of the laser, at the Larmor frequency. This rf modulation induces spin rotations about an axis in the $F^x-F^y$ plane. Since there is no prior phase reference, we define the rotation to be around $F^y$ so that a $\pi/2$ pulse maps $F^x$ onto a population difference between Zeeman states.

To avoid differential evolution of the spinor phase $\phi$, we typically perform global Raman rotations by simultaneously addressing all four ensembles (except for the direct measurement of the nullifiers described in Sec.~\ref{sec:supp_direct_null}). In this setting, we achieve a global Rabi frequency of $\Omega_\text{Raman} = 2\pi\times 12.5~$kHz.

\subsection{Readout and Fluorescence Imaging}
\label{sec:readout_fluorescence}
We characterize the multimode entangled states in our experiment by state-sensitive fluorescence imaging.  To read out a specified quadrature in the $x-p$ plane (where $x \propto F^x$ and $p \propto Q^{yz}$), we first perform a global spinor rotation by a variable angle $\phi$ and subsequently perform a $90\degr$ spin rotation about $F^y$ to convert $F^x$ to $F^z$.  The implementations of these rotations are described in Sec.~\ref{sec:supp_spin_rots}.
To ensure that the rotation angle stays close to $90\degr$ during the whole duration of our experiments, we calibrate the frequency of the Rabi oscillation every hour.

For the data shown in Figs.~\ref{fig:2_global} and \ref{fig:two_cluster} of the main text, where each subsystem (left and right) consists of two atomic ensembles, we modify the readout to minimize the impact of global technical fluctuations. Specifically, we apply a local $180\degr$ rotation about one of the ensembles in each subsystem prior to the final spin rotation, thereby mapping the symmetric mode onto one that involves a differential measurement of $F^z$ between ensembles.  Similarly, the anti-symmetric mode is mapped onto a mode that remains robust against technical noise.

To measure the atomic state populations, we collect a sequence of four images, with one detecting any population in the $f=2$ hyperfine manifold and the remaining three images detecting the populations in the three magnetic substates $m = 0, \pm 1$ within the $f=1$ manifold.  For this portion of the experimental sequence, we lower the power of the 1560~nm trapping laser to reduce the ac Stark shift of the electronically excited 5P$_{3/2}$ state and reconfine the atoms in the microtraps.
We apply two counter-propagating laser beams resonant with the $f=2\rightarrow f'=3$ transition of the $\mathrm{D}_2$ line and collect the resulting fluorescence signal on an EMCCD camera. To avoid interference of the two imaging beams, we switch them on one at a time  for 3~$\mu$s each and alternate between the two beams for 126~$\mu$s per image.
After this time, most of the atoms in $f=2$ have escaped the trapping potential due to heating, and we switch on one of the imaging beams for 150~$\mu$s to remove any residual atoms in $f=2$. To measure the atoms in the remaining states, we transfer the population in the desired state to the $f=2$ manifold via microwave pulses and repeat the imaging sequence above. To reduce the sensitivity of this transfer to magnetic field noise and microwave power fluctuations, we use a composite pulse that involves a sequence of four microwave pulses with different relative phases~\cite{levitt1986composite}.

\label{sec:imaging_calibration}

To calibrate the conversion from fluorescence signal to atom number, we employ a measurement of the atomic projection noise.  We prepare $N$ atoms in a superposition of $m=\pm 1$ by initializing all atoms in $m=0$ and then rotating by $90\degr$ about $F^y$. To isolate the projection noise, we vary the atom number $N = N_{+1} + N_{-1}$ and measure the variance of the population difference $N_{+1} - N_{-1}$. Extended Data Figure~\ref{supp_fig:ShotNoise} shows these data in units of camera counts for each of the four collective modes. For each mode, we perform a polynomial fit,
\begin{equation}
    \Var{c_{+1} - c_{-1}} = a_0 + a_1 \avg{c_{+1} + c_{-1}} +  a_2 \avg{c_{+1} + c_{-1}}^2,
\label{eq:shot_noise}
\end{equation}
where $c_m$ denotes the signal from atoms in state $m$.  The linear coefficient $a_1 = r+g$ includes the count-to-atom conversion factor $r$ and an additional contribution $g\ll r$ from photon shot noise, augmented by the excess noise of the EMCCD.  From the fit value $a_1=415\pm6$ and an independent calibration of $g=20$, we obtain the conversion factor $r = 395\pm6$ counts/atom.  This calibration is consistent with an independent measurement of the dispersive cavity shift $\delta_N = 4N\Omega$ due to $N$ atoms.

The fit offset $a_0$ and quadratic component $a_2$ provide information, respectively, about the imaging noise floor and technical noise in the fluorescence readout. The quadratic component of the fits in Extended~Data~Fig.~\ref{supp_fig:ShotNoise} determines the atom number $N\sim 1/a_2$ at which technical fluctuations become comparable to the projection noise.  For the mode with the highest technical fluctuations, we find a quadratic component $a_2 = 5\times 10^{-5}$. We therefore limit the maximal atom number in the experiment to $N \lesssim 2\times 10^4$ to ensure that projection noise dominates over technical fluctuations.  For our typical atom numbers, the background noise $a_0$ is small compared with the photon shot noise, the latter being equivalent to a fraction $g/r\approx0.05$ of the atomic projection noise.  For the direct measurement of the nullifiers in Fig.~\ref{fig:cluster_state} and the EPR steering, we subtract the photon shot noise contribution from the measured variances.  

\subsection{Measurement of Contrast $C$}\label{sec:contrast}

To compute the normalized variance $\zeta^2=\Var{F^x}/(CN)$, we extract the contrast $C$ from the same data set as the variance of $F^x$. Specifically, in terms of the Zeeman state populations $N’_m$ after the readout spin rotation, we measure both the spin component $F^x = N'_{+1}-N'_{-1}$ and the quadrupole moment $Q^{xx} = 2 (N'_{+1}+N'_{-1}-2N'_0)/3$. The quadrupole moment $Q^{xx}$ is directly proportional to the contrast $C$ in our Larmor-invariant system.

In the most general case, the contrast $C$ may be expressed exactly in terms of the collective quadrupole moments as
\begin{equation}
    C = \frac{|\avg{[F^x,Q^{yz}]}|}{2N} = \frac{|\avg{Q^{zz}}-\avg{Q^{yy}}|}{2N}. 
\end{equation}
Due to the Larmor invariance of the states generated in our experiment~\cite{SM}, $\avg{Q^{xx}}=\avg{Q^{yy}}$. Further, the three quadrupole moments sum to zero, $Q^{xx}+Q^{yy}+Q^{zz}=0$, as can be seen from the definitions in Sec.~\ref{sec:definition}.

We can thus reexpress the contrast as 
\begin{equation}
    C = \frac{3\abs{\avg{Q^{xx}}}}{2N} = \frac{N'_{+1}+N'_{-1}-2N'_0}{N}.
\end{equation}
We use this expression to normalize all variances reported in the main text.

\subsection{Steering Criterion} 
\label{sec:SteeringCriterion}
To confirm Einstein-Podolsky-Rosen (EPR) steering, we show that a measurement on the right subsystem can be used to infer the measurement results in the left subsystem with a higher precision than permitted by the local Heisenberg uncertainty relation.  To calculate the error of the inference of an observable $O$ of the left subsystem conditioned on measurements of the right subsystem, we find weights $g_i$ that minimize the conditional variance 
\begin{equation}
\Var{O_L|O_R} = \Var{O_L - \sum_{i\in{R}} g_i O_i},
\end{equation}
where $i$ indexes ensembles within the right subsystem and the weights $g_i$ capture inhomogeneities in coupling for different ensembles.  For the EPR-steered state, these variances are minimized for the $x'$ and $p'$ observables.  We measure EPR steering in both directions, requiring inferences in two directions and two quadratures.  The values of all of the conditional variances, after subtracting a small photon shot noise contribution as calibrated in Sec.~\ref{sec:readout_fluorescence}, are summarized in Extended Data Table~\ref{tab:supp_steering}. We also report the optimal values of $g_i$ for each inference. For most of the inferences, higher weight is given to the ensemble closest to the center of the cavity, which we attribute to the difference in atom-light coupling for different ensembles. 

\subsection{Graph State Generation}
\label{sec:supp_cluster}

Our prescription for preparing graph states rests on diagonalizing the adjacency matrix $A$, with the resulting eigenvectors specifying collective modes to squeeze and the eigenvalues specifying the squeezed quadratures.  Formally, the diagonal matrix $\Lambda$ of eigenvalues $\lambda_m$ is related to the adjacency matrix $A$ by
\begin{equation}
    A = V^{-1} \Lambda V,
\end{equation}
where the columns of $V$ represent the eigenmodes.  In terms of the quadrupole operators $\x = (x_1, \dots, x_M)$ and $\p = (p_1, \dots , p_M)$ on the individual sites, each eigenmode is parameterized by collective quadrature operators $\tilde{\x} = V \x$ and $\tilde{\p} = V \p$.  Re-expressing Eq.~\eqref{eq:GraphState} in terms of these collective modes,
\begin{equation}\label{eq:GraphStateColl}
    \Var{\tilde{\p} - \Lambda\tilde{\x}}\rightarrow 0,
\end{equation}
shows that the antisqueezed axis for each mode lies along the line $\tilde{p}_m = \lambda_m \tilde{x}_m$.  Thus, the squeezed quadrature is oriented at a spinor phase $\phi_m = \mathrm{arccot}\,\lambda_m$.

The experimental sequence for preparing the square graph state is presented in Extended Data Fig.~\ref{fig:supp_sequence}. For this graph,
\begin{equation}
   V = \frac{1}{2}\begin{bmatrix}
       1 & 1 & 1 & 1 \\
       1 & 1 & -1 & -1 \\
       1 & -1 & -1 & 1 \\
       1 & -1 & 1 & -1 \\
        \end{bmatrix}, 
\end{equation}
where the columns have eigenvalues  $\lambda_m = (2, 0, 0, -2)$ corresponding to the angles $\phi_m = (27\degr, 90\degr, 90\degr, 153\degr)$. We choose angles $\Phi_{1,2,3}$ for the global spinor rotations so that each mode is squeezed along the appropriate axis at the end of the sequence. Schematically, we incorporate the global spinor rotations that occur during the pair creation dynamics and Larmor rotations into these angles.

Our approach of squeezing the eigenmodes of the adjacency matrix allows for generating arbitrary graph states. In the most general case, the eigenmodes may have weighted couplings to the cavity, which could be controlled via the positions or populations of the array sites, or by driving the cavity from the side with a spatially patterned field. However, even with equally weighted couplings to the cavity, a wide variety of graphs are accessible by squeezing eigenmodes of the form $V_{jm} = \exp(i\varphi_{jm})/\sqrt{M}$.  The phases $\varphi_{jm}$ can be imprinted by local spin rotations, generalizing the $180\degr$ rotations applied in this work.  For translation-invariant graphs, the eigenmodes are spin waves with $\varphi_{jm} = j(2\pi m/M)$, and a magnetic field gradient suffices to transform between them.

\subsection{Correlation Matrix Reconstruction}
\label{sec:CorrelationMatrix}
The definition of a graph state in Eq.~\eqref{eq:GraphState} considers an ideal limit of infinite squeezing. In the following, we elaborate on the definition of the adjacency matrix for realistic states with finite squeezing and show that the square graph state generated in our experiment is consistent with this definition.  The adjacency matrix that best describes a given state is the one that minimizes $\Var{\p - A \x}$, which is given by
\begin{equation}
\label{supp_eq:adjacency_covariance}
    A_{ij} = \Cov{p_i}{x_j}/\Var{x_j}.
\end{equation}
Since $A$ is necessarily symmetric we also have $A_{ij}\propto\Cov{x_i}{p_j}$. Thus, the correlations between sites in the $x$ and $p$ bases directly reveal the adjacency matrix.

To reconstruct the correlation matrices in Figures~\ref{fig:two_cluster}e and \ref{fig:cluster_state}c from measurements of the collective modes, we use a transformation of basis to express the covariance matrix in Eq.~\eqref{eq:Corr_xp} as
\begin{equation}
    \Cov{x_i}{p_j} = V^{-1}_{im}\,\Cov{\tilde{x}_m}{\tilde{p}_{m'}} V_{m'j},
\end{equation}
where we sum over the repeated indices $m$ and $m'$. The variances of $x$ and $p$ transform analogously.

In the case of equal couplings to the cavity and equal atom number in each ensemble, the eigenmodes are independent, and $\Cov{\tilde{\x}}{\tilde{\x}}$, $\Cov{\tilde{\p}}{\tilde{\p}}$, and $\Cov{\tilde{\x}}{\tilde{\p}}$ are all diagonal. We use this assumption to extract all relevant information about the state by measuring the covariance matrix
\begin{equation}
    c_m = \begin{pmatrix}
    \Var{\tilde{x}_m} & \Cov{\tilde{x}_m}{\tilde{p}_m}\\
    \Cov{\tilde{p}_m}{\tilde{x}_m} & \Var{\tilde{p}_m}
    \end{pmatrix}
\end{equation}
for each individual eigenmode. From the variances $\sq_{\text{min},m}$ and $\sq_{\text{max},m}$ in each collective mode and the orientation $\phi_m$ of the squeezed quadrature, we calculate the covariance matrix as
\begin{equation}
    c_m = R^T(\phi_m)\begin{pmatrix}
    \sq_{\mathrm{min},m} & 0 \\
    0 & \sq_{\mathrm{max},m}
    \end{pmatrix}R(\phi_m),
\end{equation}
where $R$ is a $2\times2$ rotation matrix.

\subsection{Direct Nullifier Measurements}
\label{sec:supp_direct_null}

To confirm the efficacy of our graph state generation method, we directly measure the nullifiers $\mathbf{n} = \p - A\x$ and their variances. This measurement requires simultaneously measuring some sites in the $p$ basis and others in the $x$ basis. To perform this readout for the two-mode graph state, we first apply a variable spinor rotation $\phi$ to set the measurement basis globally and then apply a $90\degr$ readout rotation about $F^y$ only to ensembles 1 and 2. This sequence maps the observable $\nObsQ_L = -x_L\cos\phi + p_L\sin\phi$ onto a population difference between Zeeman states. Subsequent evolution under the quadratic Zeeman shift thus only affects the measurement basis in ensembles 3 and 4. After a $90\degr$ rotation about the $Q^0$ axis, we apply a second Raman rotation to the remaining ensembles to enable readout of the observable $\nObsP_R = x_R\sin\phi + p_R\cos\phi$.  The results are shown in Extended~Data~Fig.~\ref{fig:supp_nullifier_epr}a, where the red/blue data points represent normalized variances of the sum/difference $\nObs_\pm = \nObsQ_L\pm\nObsP_R$.  The nullifiers are given by $n_R = p_R - x_L = \nObs_+(0\degr)$ and $n_L = p_L - x_R = \nObs_-(90\degr)$.

To extract the nullifiers for the square graph state, the procedure is the same except that we measure sites 1 and 3 in the $x$ basis (at $\phi=0$) and apply the additional 90 degree rotation about $Q^0$ to sites 2 and 4. Thus on sites 1 and 3 we read out $\nObsQ_{1,3} = -x_{1,3} \cos \phi + p_{1,3}\sin \phi$ and on sites 2 and 4 we read out $\nObsP_{2,4} = x_{2,4} \sin \phi + p_{2,4} \cos \phi$. To extract the nullifiers, we construct the following four observables,
\begin{equation}
\begin{aligned}
    \nObs_\text{1} &= \nObsQ_{1}-\nObsP_{2}-\nObsP_{4} \\
    \nObs_\text{2} &= \nObsP_{2} + \nObsQ_{1}+\nObsQ_{3} \\
    \nObs_\text{3} &= \nObsQ_{3}-\nObsP_{2}-\nObsP_{4} \\
    \nObs_\text{4} &= \nObsP_{4}+\nObsQ_{1}+\nObsQ_{3},
\end{aligned}
\end{equation}
such that $n_{1,3} = \nObs_\text{1,3}(90\degr)$ and $n_{2,4} = \nObs_\text{2,4}(0\degr)$.  The measured normalized variances as a function of the initial spinor rotation are shown in Extended~Data~Fig.~\ref{fig:supp_nullifier_epr}b. The nullifier variances reported in the main text are obtained from the data in Figs.~\ref{fig:supp_nullifier_epr} by subtracting a small detection noise contribution, as described in Sec.~\ref{sec:imaging_calibration}.

\subsection{Entanglement Detection in Graph States}
\label{sec:entanglement_detection}
We here derive the criterion for spatial entanglement in the square graph state, which uses the nullifier variances to prove that the state is not fully separable into the four individual nodes.  Specifically, we show that all separable states are subject to a lower bound on the average value
\begin{equation}
\Wnull = \frac{1}{4} \sum_{i=1}^4 v_i
\end{equation}
of the normalized variances
\begin{equation}
v_i = \frac{\Var{n_i}}{1 + \sum_j A_{ij}^2} = \frac{\Var{n_i}}{3}
\end{equation}
of the four nullifiers $n_i = p_i - \sum_j A_{ij}x_j$, where $A_{ij}$ is the adjacency matrix of the square graph state.

The density matrix for any separable state has the general form $\rho = \sum_\alpha h_\alpha \rho_\alpha$, where $\rho_\alpha = \bigotimes_{i=1}^4  \rho_{i,\alpha}$ are product states of the four ensembles and $h_\alpha$ are probabilities satisfying $\sum h_\alpha = 1$. The nullifier variances thus satisfy
\begin{equation}
\begin{aligned}\label{eq:Wnull}
\Wnull = \frac{1}{12} \sum_{i=1}^4 \Var{n_i}_\rho \geq \frac{1}{12} \sum_\alpha h_\alpha \sum_{i=1}^4 \Var{n_i}_{\rho_\alpha},
\end{aligned}
\end{equation}
where the inequality is saturated in the absence of classical correlations between the nullifiers. For any product state $\rho_\alpha$, there are no correlations between measurements on different sites. Thus, for the square graph state,
\begin{equation}
\label{eq:supp_sep_witness1}
    \begin{aligned}
        \sum_{i=1}^4 \Var{n_i}_{\rho_\alpha} &= \sum_{i=1}^4 \Var{p_i}_{\rho_\alpha}+\sum_{i,j}A_{ij}^2\Var{x_j}_{\rho_\alpha}\\
        &=\sum_{i=1}^4\Var{p_i}_{\rho_\alpha}+2\sum_{i=1}^4\Var{x_i}_{\rho_\alpha} \\
        &\geq 8 \sqrt{2}.
    \end{aligned}
\end{equation}
In the final line, we use the local Heisenberg uncertainty relation $\Var{x_i}\Var{p_i}\geq1$ to obtain a bound on the sum of variances. Substituting this bound into Eq.~\eqref{eq:Wnull} yields $\Wnull \geq 2\sqrt{2}/3 \approx 0.94$ for all separable states.

\section*{Data and materials availability:} All data are deposited in Zenodo~\cite{zenodo}.

\section*{Acknowledgments}
This work was supported by the DOE Office of Science, Office of High Energy Physics and Office of Basic Energy Sciences under Grant No. DE-SC0019174.  A.P~and E.S.C.~acknowledge support from the NSF under Grant No. PHY-1753021. We additionally acknowledge support from the National Defense Science and Engineering Graduate Fellowship (A.P.), the NSF Graduate Research Fellowship Program (E.S.C.), and the Army Research Office under grant No. W911NF2010136 (M.S.-S.). 

\section*{Author contributions} E.S.C., P.K.~and A.P.~contributed equally. E.S.C., P.K.~and A.P. performed the experiments. All authors contributed to the analysis of experimental data, development of supporting theoretical models, interpretation of results, and writing of the manuscript. 

\section*{Competing interests} The authors declare no competing interests. 

\bibliography{cluster}
\cleardoublepage

\onecolumngrid
\setcounter{table}{0}
\renewcommand{\tablename}{\textbf{Extended Data Table}}
\renewcommand{\thetable}{\textbf{\arabic{table}}}

\setcounter{figure}{0}
\renewcommand{\figurename}{\textbf{Extended Data Fig.}}
\renewcommand{\thefigure}{\textbf{\arabic{figure}}}
\renewcommand{\theHfigure}{extdata.\arabic{figure}}
\renewcommand{\theHtable}{extdata.\arabic{table}}

\section*{Extended Data}

\begin{table}[ht]
    \centering
    \begin{tabular}{|l|c|c|}
    \hline
         & Figs.~\ref{fig:2_global} and \ref{fig:two_cluster} &  Fig.~\ref{fig:cluster_state}\\\hline
    $\delta_-$ & $-2\pi\times 1.3$~MHz & $-2\pi\times 1.6$~MHz\\
     $N$ & $1.5\times10^4$ & $8\times10^3$ \\
     $\chi$ & $-2\pi\times4.3$~kHz & $-2\pi\times1.5$~kHz\\
     $\tau$ & $50~\mu$s & $100~\mu$s\\
     \hline
    \end{tabular}
    \caption{\textbf{Summary of experimental parameters for cavity mediated interactions:} detuning $\delta_-$ of cavity drive field from two-photon resonance, total atom number $N$, collective interaction strength $\chi$ and interaction time $\tau$.}
    \label{tab:figure_parameters}
\end{table}
\newpage

{\renewcommand{\arraystretch}{1.5}%
\begin{table*}[ht]
\begin{tabular}{|l|l!{\color{lightgray}\vline}l|l|l|}
\hline
Steering Direction & Error of Inference & Value  & Optimal Weights & Steering Witness \\ \hline
\multirow{2}{*}{Right $\rightarrow$ Left}&$\Var{p'_L|p'_R}$ & $0.81\pm 0.14$ & $g_3 = 0.98, g_4 = 0.78$ & \multirow{2}{*}{$0.66\pm0.18$} \\
&$\Var{x'_L|x'_R}$ & $0.81\pm 0.12$ & $g_3 = 1.11, g_4 = 0.94$ &                                \\ \arrayrulecolor{lightgray}\hline\arrayrulecolor{black}
\multirow{2}{*}{Left $\rightarrow$ Right} & $\Var{p'_R|p'_L}$ & $0.91 \pm 0.13$ & $g_1 = 0.76, g_2 = 1.17$ & \multirow{2}{*}{$0.68\pm0.18$}              \\
&$\Var{x'_R|x'_L}$ & $0.75\pm 0.13$ &$g_1 = 0.87, g_2 = 0.81$&               \\\hline                
\end{tabular}
\caption{\textbf{Summary of EPR steering values.} To measure EPR steering between different subsystems, we need to infer the value of the left subsystem in the $x'$ and $p'$ quadratures from measurements of the right subsystem, and vice versa.  Variances representing the error of each inference, and the resulting steering witnesses, are presented.}
\label{tab:supp_steering}
\end{table*}}

\newpage

\begin{figure*}[ht]
   \centering
   \includegraphics[width=0.7\textwidth]{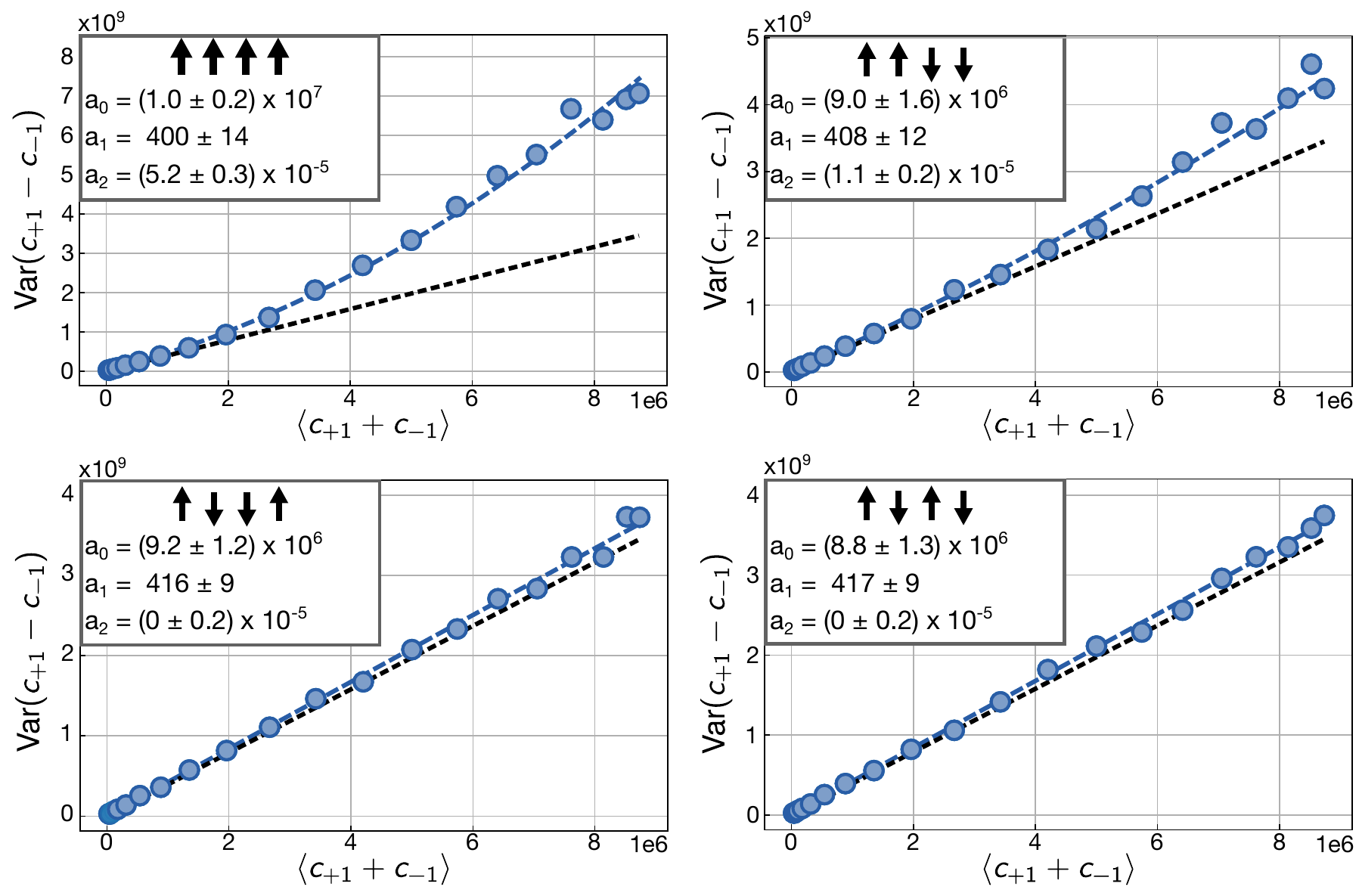}
   \caption{\textbf{Imaging calibration.} To calibrate the count-to-atom conversion factor $r$, we measure the fluctuations of the difference in counts $c_m$ from atoms in states $m=\pm1$ as a function of the average total counts $\avg{c_{+1}+c_{-1}}$ from atoms in these two states. The blue dashed line is the polynomial fit of Eq.~\ref{eq:shot_noise}, where the linear coefficient $a_1 = r+g$ accounts for the atomic projection noise and a small contribution $g\ll r$ from photon shot noise. The black dashed line represents the atomic projection noise for $r=395$~counts/atom.
   }
   \label{supp_fig:ShotNoise}
\end{figure*}

\newpage

\begin{figure}[ht]
    \centering
    \includegraphics[width=0.7\textwidth]{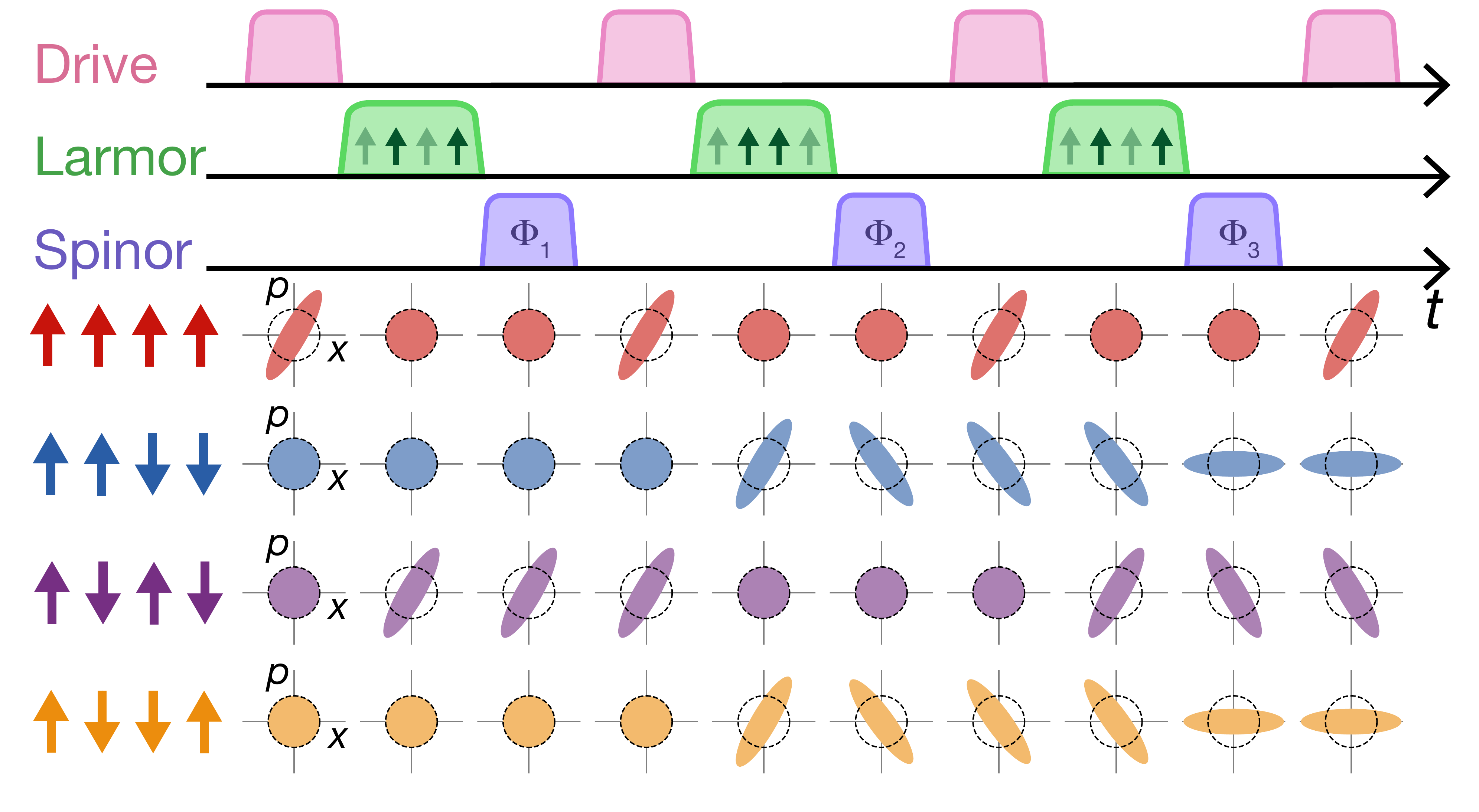}
    \caption{\textbf{Sample sequence for generating the 4-mode square graph state by squeezing collective modes.}  Bottom four rows show the state of each eigenmode throughout the entire pulse sequence. The spinor angles $\Phi_{1,2,3}= (0, 117\degr, -54\degr)$ are chosen such that, at the end of the sequence, each eigenmode is squeezed along the axis specified by the corresponding eigenvalue of the adjacency matrix.}
    \label{fig:supp_sequence}
\end{figure}

\newpage

\begin{figure}[ht]
    \centering
    \includegraphics{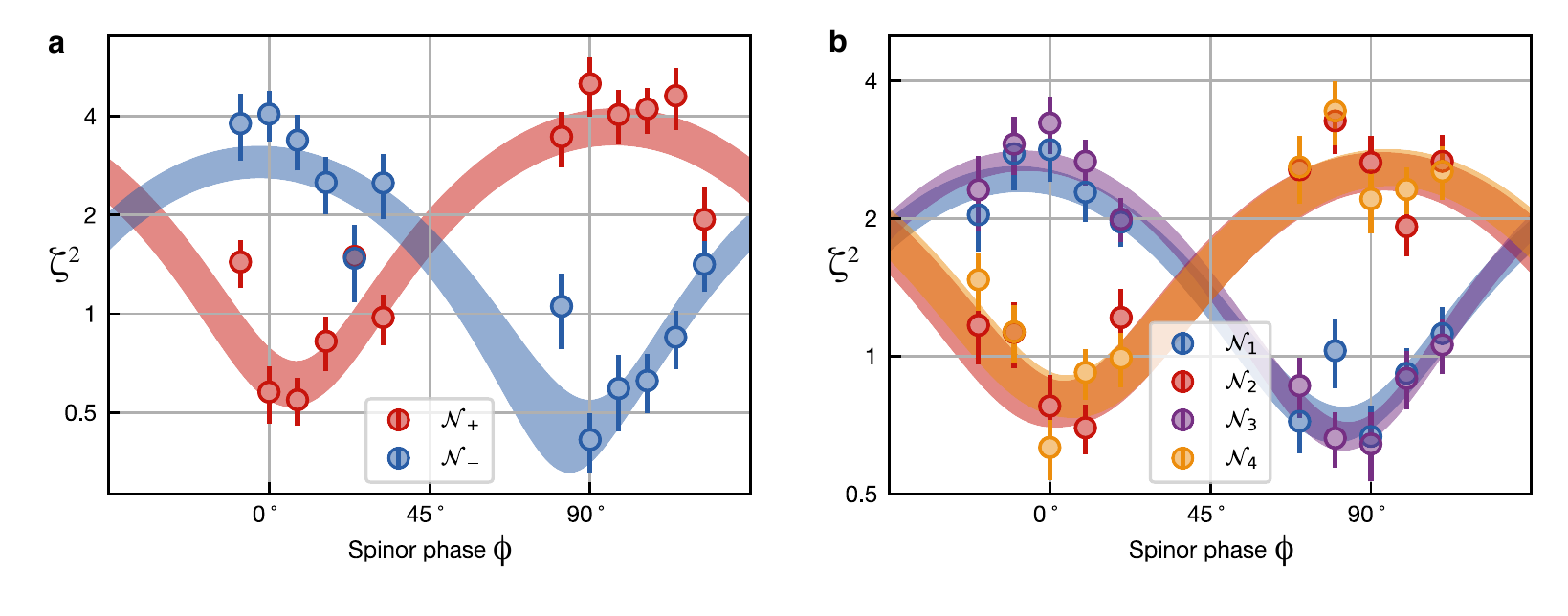}
    \caption{\textbf{Direct measurements of the nullifiers}, as described in Methods Sec.~\ref{sec:supp_direct_null}. \textbf{a}, Nullifiers for the two-mode EPR state. The spinor phase $\phi$ gives the basis in which the left ensemble pair is measured, while the right ensemble pair experiences an additional $90\degr$ of spinor evolution.  The nullifier for the left subsystem $n_L = p_L-x_R$ is extracted from $\nObs_-$ (blue) at $\phi=90\degr$.  The nullifier for the right subsystem $n_R = p_R-x_L$ is extracted from $\nObs_+$ (red) at $\phi=0\degr$. \textbf{b}, Nullifiers for square graph state. We extract the nullifier variances shown in Fig.~\ref{fig:cluster_state}c of the main text from $\nObs_{1,3}$ (blue, purple) at $\phi=90\degr$ and $\nObs_{2,4}$ (red, yellow) at $\phi = 0\degr$.}
    \label{fig:supp_nullifier_epr}
\end{figure}

\clearpage

\onecolumngrid

\renewcommand{\theequation}{S\arabic{equation}}
\setcounter{equation}{0}

\setcounter{figure}{0}
\setcounter{page}{1}
\setcounter{table}{0}
\thispagestyle{empty}

\renewcommand{\figurename}{\textbf{Fig.}}
\renewcommand{\thefigure}{S\arabic{figure}}

\renewcommand{\tablename}{\textbf{Table}}
\renewcommand{\thetable}{S\arabic{table}}
\begin{center}
	\large{\textbf{Engineering Graph States of Atomic Ensembles by Photon-Mediated Entanglement: \\ Supplementary Information}}
\end{center}
This supplement provides supporting derivations pertaining to the spin-nematic squeezing dynamics and the verification of entanglement, as well as supporting measurements. Section~\ref{sec:matrices} presents matrix representations of spin-1 operators and motivates the generalized Bloch sphere for visualization of the spin-1 dynamics.  Section~\ref{sec:metrological_squeezing} presents a derivation of the metrological squeezing parameter for spin-$f$ atoms.  Section~\ref{sec:rabi_contrast} provides an alternative determination of the contrast used in calculating normalized variances and metrological squeezing.  In Sec.~\ref{sec:squeezing_dynamics} we analytically derive the initial dynamics of spin-nematic squeezing and discuss both technical and fundamental limits on the achievable multimode squeezing.

\section{Spin-1 Algebra and Generalized Bloch Sphere}\label{sec:matrices}

The matrix forms of the spin-1 dipole operators are
\begin{equation}
	f^x = \frac{1}{\sqrt{2}}\begin{pmatrix}
		0 & 1 & 0 \\
		1 & 0 & 1 \\
		0 & 1 & 0
	\end{pmatrix},\,
	f^y = \frac{1}{\sqrt{2}}\begin{pmatrix}
		0 & -i & 0  \\
		i & 0  & -i \\
		0 & i  & 0
	\end{pmatrix},\,
	f^z = \begin{pmatrix}
		1 & 0 & 0  \\
		0 & 0 & 0  \\
		0 & 0 & -1
	\end{pmatrix}.
\end{equation}
From these matrices and the definitions in Methods Sec.~\ref{sec:definition}, we can also readily construct the matrix representations of the quadrupole operators $q^{\alpha\beta} = f^\alpha f^\beta + f^\beta f^\alpha - \frac{4}{3}I_3\delta_{\alpha\beta}$ and $q^0 = q^{zz} + \frac{1}{3}I_3$, where $I_3$ is the identity matrix and $\delta_{\alpha\beta}$ is the Kronecker delta function.  For example,
\begin{equation}
	q^{zz} = \begin{pmatrix}
		2/3 & 0    & 0   \\
		0   & -4/3 & 0   \\
		0   & 0    & 2/3
	\end{pmatrix},\,
	q^{0} = \begin{pmatrix}
		1 & 0  & 0 \\
		0 & -1 & 0 \\
		0 & 0  & 1
	\end{pmatrix},\,
	q^{yz} = \frac{i}{\sqrt{2}}\begin{pmatrix}
		0 & -1 & 0 \\
		1 & 0  & 1 \\
		0 & -1 & 0
	\end{pmatrix}.
\end{equation}

We visualize the spin-1 dynamics in an approximate $\mathrm{SU}(2)$ subspace spanned by $F^x$, $Q^{yz}$, and $Q^0$, where $F^{\alpha} = \sum_i^N f_i^{\alpha}$ and ${Q^{\alpha\beta} = \sum_i^N q_i^{\alpha\beta}}$ denote collective observables for the system of $N$ atoms. This visualization is motivated by the commutation relations
\begin{equation}
	\begin{aligned}
		[Q^{yz}, Q^{0}] & = 2i F^x              \\
		[Q^0, F^x]      & = 2iQ^{yz}            \\
		[F^x, Q^{yz}]   & = 2i(Q^0 - \Qoffset),
	\end{aligned}
\end{equation}
where $\Qoffset = (Q^{zz}+Q^{yy}-Q^{xx})/4+N/3$ commutes with $F^x$, $Q^{yz}$, and $Q^0$.  For a Larmor-invariant system with only a small side-mode population, $\avg{\Qoffset} = (N_{+1} + N_{-1})/2 \ll \avg{Q^0}$. This justifies the use of $Q^0$, which is directly accessible from the atomic populations $N_m$, as an approximation for the exact $\mathrm{SU}(2)$ subspace $\{F^x, Q^{yz}, Q^0-\Qoffset\}$. Identical dynamics occur in the subspace spanned by $F^y$, $Q^{xz}$, and $Q^0$---where the analogous approximation applies---as our system is Larmor invariant.

\section{Entanglement Detection via Metrological Gain}
\label{sec:metrological_squeezing}

Squeezing within a single mode can be used to directly witness entanglement when the generated states circumvent limits on the metrological sensitivity of unentangled states. The appropriate measure of squeezing for this context is the Wineland parameter $\xi^2$ defined in the main text, which quantifies enhanced sensitivity to rotations~\cite{wineland1994squeezed,pezze2018quantum} --- in our case on the quadrupole spin sphere with axes $\{F^x, Q^{yz}, Q^0\}$.  Entanglement is necessary for reaching values of this metrological squeezing parameter that are below the standard quantum limit (SQL) $\xi^2 = 1$.

As the seminal work defining the Wineland squeezing parameter focused on ensembles of spin-$1/2$ particles~\cite{wineland1994squeezed}, we show here how the connection between metrological gain and entanglement generalizes to systems with larger internal spin $f$, including the spin-1 atoms in this work~\cite{zou2018beating}.  Generically, the derivation of a Wineland criterion for entanglement follows from considering the metrological task of estimating a parameter $\lambda$ in a unitary transformation of the form $U = e^{i\lambda \Hpert}$.  To derive a criterion suitable for detecting spin-nematic squeezing, we assume without loss of generality that the squeezed quadrature is $F^x$ and consider the resulting enhancement in sensitivity to rotations generated by $\Hpert = Q^{yz}$.

We first derive the standard quantum limit on the sensitivity attainable to rotations about $Q^{yz}$ using an unentangled state of $N$ atoms.  For any initial state specified by a density matrix $\rho$, the quantum Cramer-Rao bound limits the precision of estimating the angle $\lambda$ in a single trial to
\begin{equation}
	\Delta\lambda \ge \frac{1}{\sqrt{\QFI[\rho,\Hpert]}},
\end{equation}
where
\begin{equation}
	\QFI[\rho,\Hpert] \leq 4\Var{\Hpert}_\rho
	\label{eq:QFI}
\end{equation}
is the quantum Fisher information~\cite{braunstein1994statistical,giovannetti2011advances}, and the inequality in Eq.~\eqref{eq:QFI} is saturated by pure states.  For product states, the quantum Fisher information for detecting rotations about $Q^{yz}$ is limited to
\begin{equation}
	F_Q[\rho,Q^{yz}] \leq 4\Var{Q^{yz}}_\rho = 4\sum_i \Var{q^{yz}_i}_\rho \leq 4 N f^2,
\end{equation}
where in the last inequality the single-atom spin $f=1$ dictates that $\Var{q^{yz}_i} \leq f^2$.  Thus, the minimum imprecision of a measurement of $\lambda$ using an unentangled state is $\Delta\lambda_\mathrm{SQL} = 1/(2f\sqrt{N})$.

The Wineland parameter compares the sensitivity of an arbitrary state to the standard quantum limit.  Assuming that the state is used to estimate $\lambda$ via the dependence of $\avg{F^x}$ on the rotation about $Q^{yz}$, the uncertainty $\Delta \lambda$ is related to the spin variance $\Var{F^x}$ by
\begin{equation}
	\left(\Delta \lambda\right)^2 = \frac{\Var{F^x}}{\left(d\avg{F^x}/d\lambda\right)^2},
\end{equation}
where the rate of change $d\avg{F^x}/d\lambda$ of $\avg{F^x}$ under the Hamiltonian $\Hpert = Q^{yz}$ is given by
\begin{equation}
	\abs{\frac{d\avg{F^x}}{d\lambda}} = \abs{\avg{[F^x,Q^{yz}]}}.
\end{equation}
The ratio of the measurement variance $(\Delta\lambda)^2$ to the standard quantum limit $(\Delta\lambda_\mathrm{SQL})^2$ is thus
\begin{equation}\label{eq:Wineland}
	\xi^2 \equiv \left(\frac{\Delta \lambda}{\Delta \lambda_\mathrm{SQL}}\right)^2 = \frac{4f^2 N\Var{F^x}}{\abs{\avg{[F^x,Q^{yz}]}}^2} = \frac{f^2\Var{F^x}}{C^2 N}.
\end{equation}
For our atoms with internal spin $f=1$, Eq.~\eqref{eq:Wineland} simplifies to $\xi^2 = \Var{F^x}/(C^2 N)$.  The Wineland parameter is related to the normalized variance $\sq$ plotted in the main text as $\xi^2= \sq/C$, so that the SQL is represented by a line at $\zeta^2 = C$.

\section{Calibrating Contrast and Confirming Larmor Invariance}
\label{sec:rabi_contrast}

To calculate the normalized variances presented in the main text, we determine the commutator $CN$ from the average populations of the Zeeman states after the readout spin rotation, as described in the Methods (Sec. \ref{sec:contrast}).  The ability to determine both the noise $\Var{F_x}$ and the commutator $CN=\abs{\avg{[F^x,Q^{yz}]}}/2$ from the same data set is a feature of the Larmor-invariant spin-1 system and is advantageous for minimizing sensitivity to any drifts in experimental parameters.  However, an alternative approach is to determine the contrast $C$ by a separate interferometric measurement, as suggested by the role of the metrological squeezing parameter in quantifying enhanced interferometric sensitivity.  Here, we present such a measurement of the contrast $C_\text{Rabi}$ of a Rabi oscillation on the $\{F_x,Q_{yz},Q_0\}$ sphere, which corroborates the method used in the main text and confirms the assumption of Larmor invariance.

\begin{figure}
	\centering
	\includegraphics[width=\textwidth]{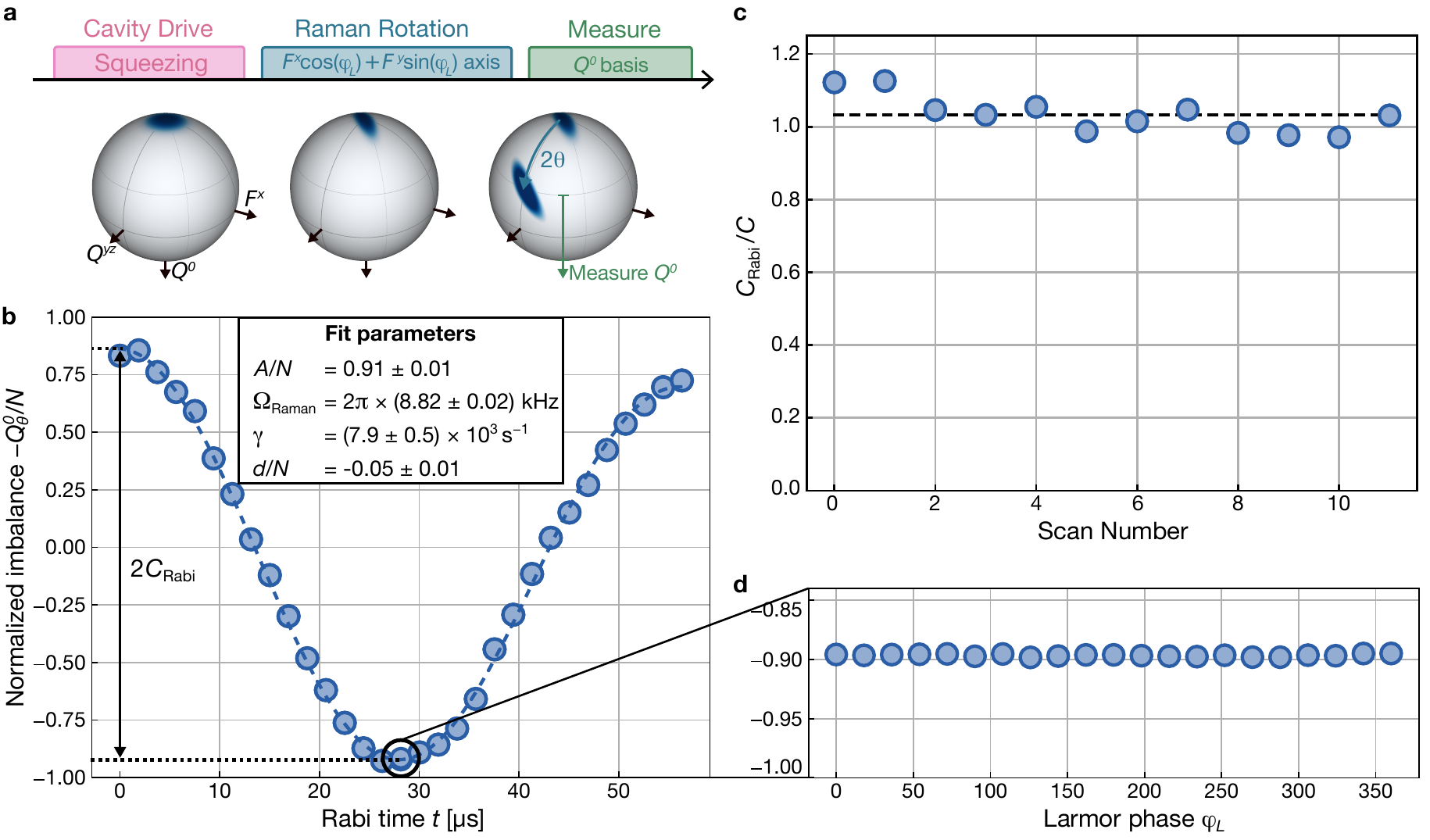}
	\caption{\textbf{Direct measurement of contrast.} \textbf{a}, 
	After applying the cavity drive field for 50\,$\micro{s}$ to generate spin-nematic squeezing, we perform a rotation about an arbitrary axis with Larmor phase $\varphi_L$ in $F^x-F^y$ plane. When $\varphi_L =0$ this rotation corresponds to Rabi oscillations that are visualized on the $\{F^x, Q^{yz}, Q^0\}$ sphere. 
	\textbf{b}, Following the rotation, we measure the normalized imbalance $-Q^0_\theta/N = (N'_0-(N'_{+1}+N'_{-1}))/N$ as a function of Rabi oscillation time $t$. The dashed line is a fit of the model in Eq.~\eqref{eq:rabi_model}. 
	From this fit, we extract the contrast $C_\text{Rabi}= \abs{Q^0_{\theta=0\degr} - Q^0_{\theta=90\degr}}/(2N) = 0.89 \pm 0.01$.  
	\textbf{c}, We compare the contrast $C_{\mathrm{Rabi}}$ to the value $C = (3 Q^0_{\theta=90\degr} - 1)/(2N)$ computed as derived in Methods Sec.~\ref{sec:contrast}. 
	The percent-level difference indicated by the average ratio $C_{\text{Rabi}}/C = 1.03 \pm 0.01$ (black dashed line) is attributable to the dephasing of the Rabi oscillation at rate $\gamma$,
	 which results in conservative estimates of normalized variances and squeezing parameters in the main text.  \textbf{d},
	  Measurements of population imbalance after a rotation by $\theta = 90\degr$ confirm that the calibration of 
	  contrast is independent of the choice of Larmor phase $\varphi_L$. Data shown are the average over typically 50 iterations.
	  }
	\label{fig:supp_normcheck}
\end{figure}

We determine the contrast $C_\text{Rabi}$ from the amplitude of a Rabi oscillation after spin-nematic squeezing. For this measurement, we prepare the atoms in $m_F=0$ before driving the cavity for 50~$\micro{s}$ to induce pair-creation dynamics. Up to this point, the sequence is the same as that used for the data in Fig.~\ref{fig:2_global} of the main text. After the dynamics, we apply a spin rotation by an angle $\theta = \Omega_\text{Raman} t$. Assuming Larmor invariance of the state, this operation is equivalent to a rotation around $F^x$ by an angle $2\theta$ on the $\{F^x, Q^{yz}, Q^0\}$ sphere (Fig.~\ref{fig:supp_normcheck}a). We then evaluate the normalized population imbalance $-Q^0_\theta/N \equiv (N_0' - (N_{+1}'+N_{-1}'))/N$ as function of $\theta$. The full expression for the resulting Rabi oscillation in terms of the initial expectation values $\avg{Q^0}$, $\avg{Q^{yz}}$, and $\avg{\Qoffset} = \avg{Q^{zz}+Q^{yy}-Q^{xx}}/4+N/3$ is
\begin{equation}\label{eq:Rabi}
	Q^0_\theta = \cos(2\theta) \avg{Q^0} + \sin(2\theta) \avg{Q^{yz}} + (1-\cos(2\theta))\avg{\Qoffset}.
\end{equation}
The final term arises from the fact that $\{F_x,Q_{yz},Q_0\}$ form an approximate rather than exact $\mathrm{SU}(2)$ subspace (see Sec.~\ref{sec:matrices}) and is small, $\avg{D} \approx (N_{+1}+N_{-1})/2\ll\avg{Q^0}$, for the early-time squeezing dynamics in our system.  The contrast of the Rabi oscillation can be expressed in terms of the operators prior to the spin rotation as
\begin{equation}
	\begin{aligned}
		C_\text{Rabi}
		 & = \frac{1}{2N}\abs{Q^0_{\theta=0\degr} - Q^0_{\theta=90\degr}} \\
		 & = \frac{1}{2N}\left|2\avg{Q^0-\Qoffset}\right|                 \\
		 & =\frac{1}{2N} \left|\avg{[F^x, Q^{yz}]}\right|,
	\end{aligned}
\end{equation}
which provides an alternative measurement of the contrast $C$ as defined in the main text.

To extract the contrast $C_\text{Rabi}$, we fit measurements of the population imbalance $Q^0(t)$ after a Raman pulse of duration $t$ with a function of the form
\begin{equation}\label{eq:rabi_model}
	-Q^0(t) = A\, e^{-(\gamma \,t)^2}\,\cos(2\Omega_\text{Raman} t)  + d,
\end{equation}
where $\Omega_\text{Raman}$ is the Rabi frequency and $\gamma$ is the Rabi dephasing rate. The amplitude $A$ and offset $d$ together account for both the final term in Eq.~\eqref{eq:Rabi} and incoherent collisional spin-exchange interactions between the readout rotation and imaging, where the latter is dominant in our system. We use the fit to extract the contrast $C_\text{Rabi} = \abs{Q^0(t =\pi/2\Omega_\text{Raman}) - Q^0(t=0)}/(2N) = 0.89(1)$.

The primary source of imperfect contrast is spin-changing collisions that occur in the time (20~ms) between the Raman rotation and the fluorescence readout. These collisions permit states polarized with either all atoms in $m=0$ or all atoms in $m=\pm1$ to decay into a mixture of all three states. This process reduces contrast and introduces an offset to the Rabi oscillation, but does not directly affect the measurement of $\Var{F^x}$, since the collisions preserve the population difference $N'_{+1} - N'_{-1}$ from which we infer the polarization $F^x$ in the experiment. Additionally, dephasing from inhomogeneity in the Raman rotation and detuning due to a finite ratio $\Omega_\mathrm{Raman}/q$ reduce the measured value of $\abs{Q^0_{\theta=90\degr}}$.  The measurement of contrast $C$ used for evaluating normalized variances depends only on $Q^0_{\theta=90\degr}$ and not on $Q^0_{\theta=0\degr}$, and is thus more strongly impacted by these effects, leading to a $3\%$ difference between the two measurements as shown in Fig.~\ref{fig:supp_normcheck}c.  This difference indicates that the normalized variances and squeezing parameters reported in the main text are conservative estimates.

In calibrating the contrast $C$ via Rabi oscillations, we perform the Raman rotation about an arbitrary axis in the $F^x-F^y$ plane, as there is no prior phase reference in the experiment.  We thus implicitly assume Larmor invariance of the generated state.  We confirm this assumption by performing repeated measurements of the population imbalance at fixed rotation angle $\theta$ while varying the Larmor phase $\varphi_L$. We perform these measurements with a readout rotation of $\theta = 90\degr$ where the measured imbalance is most sensitive to changes in the Larmor phase, and we observe no dependence on phase over a full Larmor oscillation period (see Fig.~\ref{fig:supp_normcheck}d).

\section{Squeezing Dynamics}
\label{sec:squeezing_dynamics}
We present a theoretical formulation of the spin-nematic squeezing dynamics. This model provides a basis to estimate the contributions of technical noise in our measured squeezing, as well as the effects of cavity dissipation. Finally, we discuss fundamental limits set by the cavity cooperativity on the degree of multimode squeezing attainable in our protocol for preparing graph states.

\subsection{Equations of Motion}\label{sec:eom}
We analyze the dynamics of spin-nematic squeezing for a system initialized with all atoms in $m=0$, focusing on the experimentally relevant regime of early times where the population in states $m=\pm 1$ remains small ($N_{+1} + N_{-1} \ll N$).  To allow for effects of finite atomic temperature, we begin by writing down a Hamiltonian that incorporates non-uniform coupling to the cavity mode,
\begin{equation}
	H = \frac{\chi}{2N} \left(\mathcal{F}^x\mathcal{F}^x+\mathcal{F}^y\mathcal{F}^y\right) + \frac{q}{2} Q^{0}.
\end{equation}
Here the collective spin $\mathbf{F}$ defined in the main text is replaced with a weighted collective spin $\boldsymbol{\mathcal{F}} = \sum_i w_i f_i$, which includes a correction for inhomogeneous cavity couplings $w_i \propto \Omega_i$, where $\Omega_i$ denotes the ac Stark shift per intracavity photon experienced by the $i^\mathrm{th}$ atom.  The weights $w_i$ are normalized such that $\avg{w_i}=1$.

We describe the early-time dynamics in the two dimensional subspace spanned by the weighted spin operators $\mathcal{F}^x$ and $\mathcal{Q}^{yz}$. During these early times, commutators relevant to the dynamics are $[\curlyF, \mathcal{F}^y]=2i\sum_i w_i^2 f_i^z\approx 0$, $[\curlyF,Q^{yz}]\approx -2iN$, and $[Q^0,\curlyF]=2i\mathcal{Q}^{yz}$. The Heisenberg equations $d\mathcal{O}/dt = i [H,\mathcal{O}]$ for both spin observables in this space are
\begin{equation}
	\begin{aligned}
		\frac{d}{dt}\begin{bmatrix}
			            \mathcal{F}^x \\ \mathcal{Q}^{yz}
		            \end{bmatrix} & = \left[
			\begin{array}{cc}
				0        & -q \\
				q+2 \chi & 0  \\
			\end{array}\right] \begin{bmatrix}
			                   \mathcal{F}^x \\ \mathcal{Q}^{yz}
		                   \end{bmatrix}.
		\label{eq:HamiltonianDynamics}
	\end{aligned}
\end{equation}
Identical dynamics occur in the subspace spanned by $\mathcal{F}^y$ and $\mathcal{Q}^{xz}$ so that state remains invariant under global spin rotations about $F^z$.

The linear equations of motion for $\mathcal{F}^x$ and $\mathcal{Q}^{yz}$ can be solved exactly. This system has eigenvalues $\pm \lambda$ where $\lambda = \sqrt{-q (q+2 \chi )}$. The corresponding solutions are
\begin{equation}
	\begin{aligned}
		\mathcal{F}_{+\lambda}(t) & = \frac{e^{\lambda t}}{\sqrt{1+(\lambda/q)^2}} \left(\mathcal{F}^x(0) - \frac{\lambda}{q} \mathcal{Q}^{yz}(0)\right)   \\
		\mathcal{F}_{-\lambda}(t) & = \frac{e^{-\lambda t}}{\sqrt{1+(\lambda/q)^2}} \left(\mathcal{F}^x(0) + \frac{\lambda}{q} \mathcal{Q}^{yz}(0)\right).
	\end{aligned}
\end{equation}
In general, these two operators are not orthogonal unless $\chi=-q$. The expectation value and variance of any observable $\mathcal{F}_\phi = \cos{\phi}\, \mathcal{F}^x - \sin{\phi}\,\mathcal{Q}^{yz}$ can be calculated from these operators by noting that  $\mathcal{F}_\phi(t) = a \mathcal{F}_{-\lambda}(t)+b\mathcal{F}_{+\lambda}(t)$ where $a$ and $b$ are real coefficients that are independent of time and may be solved for from the expressions at time $t=0$. The variance for a particular spinor angle $\phi$ at time $t$ is then given by
\begin{equation}
	\avg{\mathcal{F}_\phi(t)^2} = e^{-2\lambda t}a^2 \avg{\mathcal{F}_{-\lambda}(0)^2} + ab\avg{\{\mathcal{F}_{+\lambda}(0),\mathcal{F}_{-\lambda}(0)\}} +  e^{2\lambda t}b^2 \avg{\mathcal{F}_{+\lambda}(0)^2},
	\label{eq:supp_sqeeze_dyn}
\end{equation}
where the cross term proportional to $ab$ again highlights that $\mathcal{F}_{-\lambda}(t)$ and $\mathcal{F}_{+\lambda}(t)$ are in general not orthogonal. At $t=0$ the system is in a coherent state with variance $\avg{\mathcal{F}_\phi(0)^2}=N$ at the level of  projection-noise for all values of $\phi$. This condition yields the constraint
\begin{equation}
	a^2+b^2+2ab \frac{1-(\lambda/q)^2}{1+(\lambda/q)^2}=1,
	\label{eq:supp_constr}
\end{equation}
which reduces to $a^2+b^2=1$ when $\chi=-q$.

The operator with maximal variance (the anti-squeezed quadrature) for $t\gtrsim 1/\lambda$ is determined by maximizing the coefficient $b$ of the exponentially growing mode (see Eq.~\eqref{eq:supp_sqeeze_dyn}) subject to the constraint of Eq.~\eqref{eq:supp_constr}. This is achieved when $b = -\chi/\lambda$ and $a = -(\chi+q)/\lambda$, corresponding to an anti-squeezing $\sqmax = |\chi/\lambda|^2$ $e^{2\lambda t} $ at an angle $\phi = \arctan(\lambda/q)$. Since the dynamics preserve phase space area, this corresponds to a minimum variance $\sqmin = 1/\sqmax$ at an angle $\phi_{\mathrm{min}} = \arctan{(-q/\lambda)}$.

\subsection{Technical Limitations on Squeezing}\label{sec:technical_noise}
The model of the dynamics in Sec.~\ref{sec:eom} provides a foundation for estimating the effect of experimental noise on the amount of observed squeezing.  Two primary limits are fluctuations in the collective interaction strength $\chi$ and inhomogeneous coupling of the thermal distribution of atoms to the cavity, where the latter effect introduces a slight discrepancy between the collective mode squeezed by the cavity and the observable detected in fluorescence imaging.

\subsubsection{Fluctuations in Interaction Strength}

The collective interaction strength $\chi$ varies the angle of squeezing as $\phimin = \arctan(-q/\lambda)$, and so fluctuations in $\chi$ act to wash out the squeezing.  We can write the variance as a function of spinor phase as
\begin{equation}\label{supp_eq:sqphi}
	\sq(\phi) = \sqmin + \sin^2\left(\phi - \phimin\right)\left(\sqmax - \sqmin\right).
\end{equation}
To find the effect of fluctuations $\Delta\chi$ in interaction strength, we expand Eq.~\eqref{supp_eq:sqphi} around the angle $\phimin$ of optimum squeezing and write our expression in terms of $\chi$.
To leading order in $\Delta\chi$, we are left with the additional noise $\Delta\sq_{\mathrm{interaction}}$ from interaction strength fluctuations, which is given by
\begin{equation}
	\Delta\sq_{\mathrm{interaction}} = \sqmeas - \sqmin = \sqmax \frac{q}{2|q + 2\chi|}\left(\frac{\Delta\chi}{\chi}\right)^2.
\end{equation}
This is a lower bound on the added noise, since the angle $\phimin$ at finite times has larger fluctuations than in the $t>\lambda^{-1}$ limit. In principle, added noise from interaction strength fluctuations can be suppressed by working in the regime of $\chi\gg q$.  However, we do not operate in this limit because setting $\chi\sim q$ maximizes the squeezing rate $\lambda$ relative to the fundamental cavity dissipation, as we shall see in Sec.~\ref{sec:cavity_dissipation}.

The fluctuations in interaction strength $\chi$ in our experiment arise from variations in the number of intracavity photons $\overline{n}$, the number of coupled atoms $N$, or the detunings from the two virtual Raman processes $\delta_{\pm}$ (see Eq.~\eqref{supp_eq:chi_def}). These sources of noise are correlated, as the number of intracavity photons
\begin{equation}
	\overline{n} = \overline{n_i} \frac{(\kappa/2)^2}{\delta_c^2+(\kappa/2)^2}
\end{equation}
not only depends on the input drive strength $\overline{n_i}$, but also depends on the detuning from cavity resonance. The detunings $\delta_c$ and $\delta_{\pm}$ in turn depend on the atom number $N$ due to the dispersive shift $\delta_N = 4\Omega N$ of the cavity resonance induced by the atoms.  The two direct sources of fluctuations in $\chi$ are then the number of input photons $\overline{n_i}$ and the atom number $N$, which lead to total fluctuations
\begin{equation}
	\left(\frac{\Delta\chi}{\chi}\right)^2 =
	\left(\frac{\Delta{\overline{n_i}}}{\overline{n_i}}\right)^2 + \alpha^2\left(\frac{\Delta N}{N}\right)^2,
\end{equation}
where
\begin{equation}
	\alpha = 1 + \left|\frac{2\delta_N}{\delta_c}\right|+\left|\frac{\delta_N}{\delta_-}\right|
\end{equation}
evaluates to $\alpha\approx 2$ for our parameters.
We stabilize the drive input power to ensure $\Delta \overline{n_i}/\overline{n_i} < 5\%$, and we reduce fluctuations in atom number by post-selecting so that $\Delta N/N < 5\%$ within each data set.  At $\Delta\chi/\chi <10\%$, and a typical value of $\sqmax = 10$, using the values from Methods Sec.~\ref{supp_sec:interactions}, the additional noise from interaction strength fluctuations is $\Delta\sq_{\mathrm{interaction}} \approx 0.02$.

\subsubsection{Inhomogeneous Atom-Cavity Coupling}

Our fluorescence imaging measures a uniformly weighted collective spin $F^x$, while the cavity couples to the inhomogeneously weighted collective spin $\mathcal{F}^x$ defined in Sec.~\ref{sec:eom}.  Any width in the distribution of coupling weights $w_i$ manifests itself in reduced squeezing.  Without loss of generality, we assume $\curlyF$ is the squeezed observable and compute the projection of the measured observable on the squeezed observable: $\text{Tr}({F^x\curlyF})/(|\curlyF||F^x|) = \avg{w_i}/\sqrt{\avg{w_i^2}}$. The excess noise is given by the magnitude of the remaining component,
\begin{equation}
	\label{supp_eq:inhomogeneous_coupling}
	\Delta\sq_{\mathrm{coupling}}= 1 - \frac{\avg{w_i}^2}{\avg{w_i^2}} = \frac{\Var{w_i}}{\Var{w_i} + \avg{w_i}^2}.
\end{equation}

The variance in couplings comes primarily from the thermal distribution of the atomic states. We parameterize the temperature by the ratio $\beta = U_0/(k_B T)$ of the lattice depth to the atomic temperature. Assuming a harmonic trap, the excess noise Eq.~\eqref{supp_eq:inhomogeneous_coupling} for a single lattice site is given by
\begin{equation}
	\label{supp_eq:temperature_model}
	\Delta\sq_{\mathrm{coupling}} = 1 - \frac{2\beta(4 + \beta)}{\left(\beta + 2\right)^2\left(\exp\left(8\beta^{-1}\right)- 2\exp\left(4\beta^{-1}\right) + 3\right)}.
\end{equation}
In the low temperature limit $\beta\rightarrow\infty$, to leading order the added noise is
\begin{equation}
	\label{supp_eq:temperature_model_expansion}
	\Delta\sq_{\mathrm{coupling}} = 12 \beta^{-2}.
\end{equation}
For an ensemble near cavity center with an inverse temperature of $\beta = 15$, Eq.~\eqref{supp_eq:temperature_model} limits squeezing to $\sq > 0.08$.

\begin{figure}
	\centering
	\includegraphics[width=0.5\textwidth]{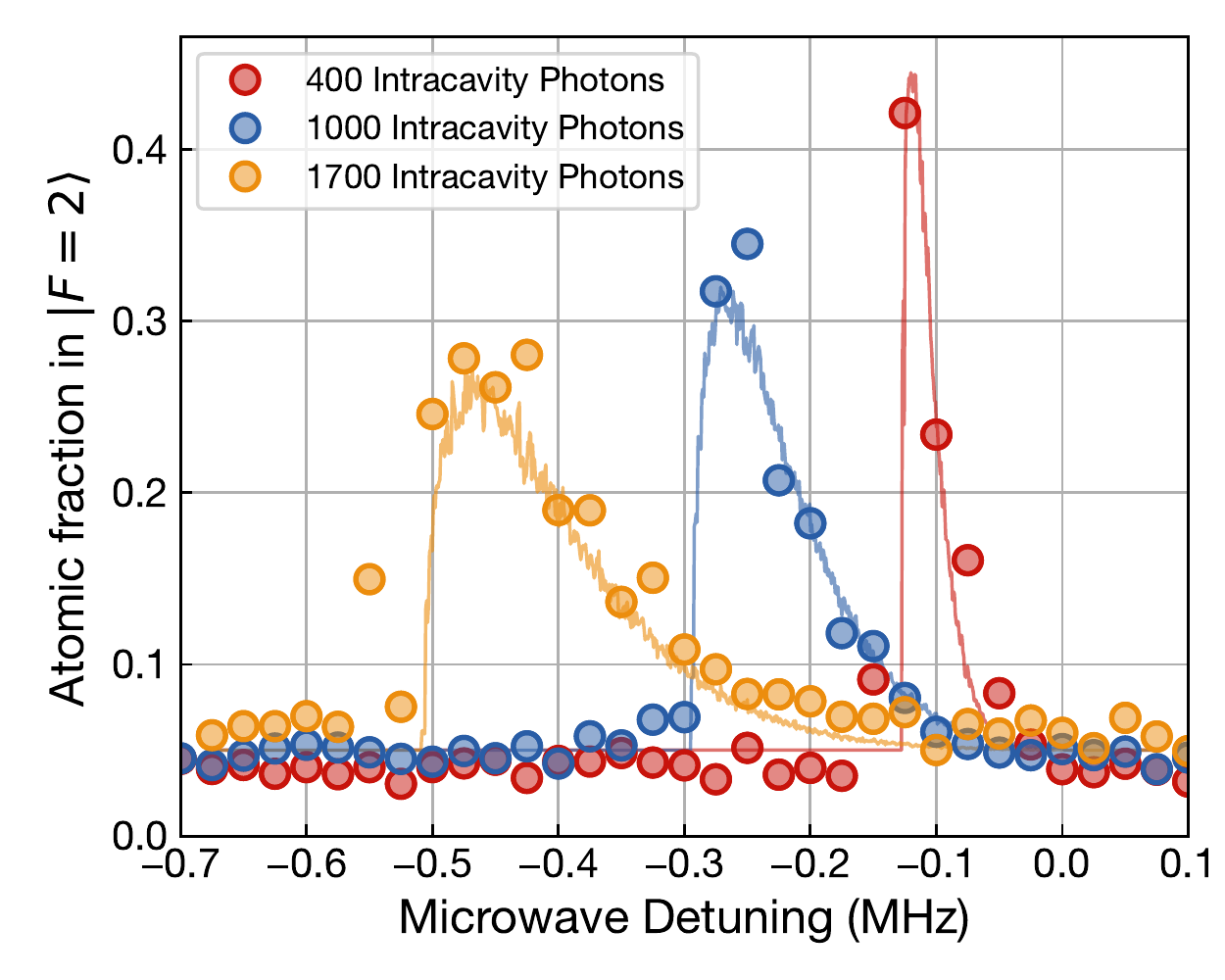}
	\caption{Measurement of the drive-induced ac Stark shift on the $\ket{f = 1, m_f = 0}\rightarrow\ket{2, 0}$ transition. Red, blue, and gold indicate increasing intracavity photon number.  The measured distributions of Stark shifts (circles) are well described by a numerical fit based on a thermal distribution of atomic positions with a ratio $\beta = 15$ of trap depth to temperature.}
	\label{supp_fig:temp_spectroscopy}
\end{figure}

We directly measure the distribution of couplings $w_i$ via microwave spectroscopy, probing the ac Stark shift induced by the drive field on the hyperfine clock transition $\ket{f = 1, m_f = 0}\rightarrow\ket{2, 0}$.  The drive light, detuned from atomic resonance by $\Delta = -2\pi\times 9.5$~GHz, induces a differential ac Stark shift that is directly proportional to the weight $w_i$ for each atom.  We measure the distribution of Stark shifts at different drive intensities, as shown in Fig.~\ref{supp_fig:temp_spectroscopy}. The measured spectra are well fit by a model of a thermal distribution with inverse temperature $\beta=15$, exhibiting variances and means that directly corroborate the bound established in Eqs.~\eqref{supp_eq:inhomogeneous_coupling}-\eqref{supp_eq:temperature_model}.

\subsection{Cavity Dissipation}
\label{sec:cavity_dissipation}

The unitary dynamics described in Methods Sec.~\ref{sec:eom} are modified by decay channels inherent to any real cavity system. As described in Sec.~\ref{supp_sec:interactions} and Refs.~\cite{davis2019photon,periwal2021programmable}, spin-exchange interactions in the cavity are mediated by a virtual process in which atoms collectively scatter photons from a vertically polarized drive field into a horizontally polarized cavity mode. For coherent interactions, these horizontally polarized photons are subsequently scattered back into the vertical drive mode, allowing for unitary transfer of information among the atoms. However, in practice, photons may also be lost before completing the unitary dynamics and thereby carry away quantum information.

A photon may be lost either due to the finite cavity lifetime or by atomic scattering into free space. In the case of cavity decay, the loss of a photon is accompanied by creation or annihilation of a collective spin excitation.  This decay channel is described by the Lindblad operators $L_\pm = \sqrt{\gamma_\pm} \mathcal{F}^\pm$ and has a characteristic strength $\Gammac= 2N(\gamma_++\gamma_-)$ that, in analogy to the collective interaction strength $\chi$, is enhanced by the number of atoms. The ratio of the collective decay to the collective interaction strength
\begin{equation}
	\frac{\Gammac}{\chi} = \frac{\kappa}{\delta_-}
\end{equation}
is determined by the detuning $\delta_-$ from the dominant Raman process in our experiment.

To quantify the impact of the collective decay process on the squeezing, we write down the Lindblad equation of motion for the squeezed quadrature $\mathcal{F}_{-\lambda}$,
\begin{equation}
	\frac{d \avg{\mathcal{F}_{-\lambda}^2}}{dt} = \avg{ i [H, \mathcal{F}_{-\lambda}^2] + \sum_r 1/2 \left(L_r^\dagger [\mathcal{F}_{-\lambda}^2, L_r] + [L_r^\dagger,  \mathcal{F}_{-\lambda}^2]L_r\right)},
\end{equation}
with the two loss operators $L_\pm = \sqrt{\gamma_\pm} \mathcal{F}^\pm$. Under the simplifying assumptions of uniform couplings ($\mathcal{F^\pm} = F^\pm$), $\chi \approx -q$ and sufficiently early-time dynamics, the equation of motion for squeezing is
\begin{equation}
	\frac{d \avg{\mathcal{F}_{-\lambda}^2}}{dt} = -2\lambda \avg{\mathcal{F}_{-\lambda}^2} + N\Gammac.
	\label{eq:dsqdt_2ph}
\end{equation}
The steady state of this equation leads to a bound on the variance due to collective decay of
\begin{equation}
	\label{eq:cavity_collective}
	\Delta\sq_\text{coll}= \avg{\mathcal{F}_{-\lambda}^2}/N \geq \Gammac/(2\lambda).
\end{equation} The variance exponentially decays to this bound at a rate proportional to $\exp(-2\lambda)$. At finite times, the effect of the bound is mathematically equivalent to mixing the ideal squeezed state achieved under unitary dynamics with vacuum fluctuations on a beam splitter with transmission $1-\Delta\sq_\text{coll}$.

Additionally, photons may be lost due to free-space scattering at a rate $\Gamma_\text{sc}$ per atom. Free-space scattering is not a collective process, as the scattered photons carry away information about individual atoms.  On cavity resonance, free-space scattering is thus suppressed with respect to interactions by a factor $N\eta/k$, where $N\eta$ is the collective cooperativity for a cycling transition and the numerical factor $k=96$ includes the strengths of the atomic transitions in our level scheme~\cite{periwal2021programmable}. Overall, the rate of free-space scattering in the limit $\delta_- > \kappa$ is then
\begin{equation}
	\frac{\Gamma_\text{sc}}{\chi} = \frac{96}{N\eta} \frac{\delta_-}{\kappa}.
	\label{eq:cavity_loss_params}
\end{equation}

The effect of a scattering event is to erase correlations between the atom that scatters a photon and the remaining atoms.  This erasure of correlations adds noise to the squeezed quadrature at a rate
\begin{equation}
	\label{eq:cavity_scattering}
	\frac{d\avg{\mathcal{F}^2_{-\lambda}}}{dt} = \alpha N \Gamma_{\text{sc}},
\end{equation}
where $\alpha=2$ for the worst case where an atom is projected into the $m=0$ state. However, while collective decay only impacts the mode coupled to the cavity, free-space scattering continues to impact all modes that have already been squeezed. On average each mode is impacted by scattering for a total duration $\tau\times M/2$, where $\tau$ is the duration of each squeezing pulse, yielding a noise contribution
\begin{equation}
	\Delta\zeta_{\mathrm{sc}}^2 = M\tau\Gamma_{\text{sc}}.
\end{equation}
While free-space scattering additionally reduces the contrast by a factor of approximately $e^{-M\tau\Gsc/2}$, in the regime of strong squeezing the added noise is the dominant limitation.

The impact of the two noise contributions $\Delta\sq_\text{coll}$ and $\Delta\zeta_{\mathrm{sc}}^2$ can be minimized by choosing optimal values of the interaction strength $\chi$ and detuning $\delta_-$. An interaction strength of $\chi=-q$ optimizes the speed of the coherent dynamics $\lambda/\chi$.  Optimizing the two limits for the mode with maximal scattering, with $\tau\lambda = 1$, yields a detuning of $\delta_- = \kappa\sqrt{N\eta/(192M)}$, balancing the impact of the two loss channels to minimize their combined effect. In practice, we experimentally optimize squeezing, which takes into account additional noise sources, resulting in a slight deviation from the theoretical optimum.

Having derived expressions for both collective decay and free-space scattering, we summarize the impact on the experiments in the current work in Sec.~\ref{sec:noise_summary} and derive fundamental limits on the scaling of the squeezing in Sec.~\ref{sec:fundemental_scaling}.

\subsection{Summary of Noise Contributions}\label{sec:noise_summary}
We summarize the impact of all noise processes limiting our squeezing in Table~\ref{tab:squeezing_limits}. The effects of cavity decay, free-space scattering, and coupling variation are all mathematically equivalent to mixing the squeezed quantum state with zero point fluctuations, as if on a beam splitter. Starting from the minimal possible variance given unitary dynamics, $1/\sqmax$, each process results in a factor of $(1-\Delta\sq_i)$ reduction in the amount by which the state is squeezed. We calculate the combined effect of these noise sources $\Delta\sq_i$, assuming that they are all independent, as
\begin{equation}
	\Delta\sq_{\mathrm{total}} = 1 - \Pi_{i}(1 - \Delta\sq_i).
\end{equation}
Photon shot noise and interaction strength noise behave differently, since they directly add noise rather than degrading the state toward the standard quantum limit. These terms are added at the end using standard propagation of uncertainty. Finally, we divide by the Rabi oscillation contrast $C_\mathrm{Rabi} = 0.89$ measured in Sec.~\ref{sec:rabi_contrast} to obtain the expected normalized variance $\zeta^2$.

\begin{table}[]
	\centering
	\begin{tabular}{|l|c|c|}
		\hline
		                                  & Figs.~\ref{fig:2_global} and \ref{fig:two_cluster} & Fig.~\ref{fig:cluster_state} \\\hline
		$\delta_-$                        & $-2\pi\times 1.3$~MHz                              & $-2\pi\times 1.6$~MHz        \\
		$N$                               & $1.5\times10^4$                                    & $8\times10^3$                \\
		$\chi$                            & $-2\pi\times4.3$~kHz                               & $-2\pi\times1.5$~kHz         \\
		$\lambda$                         & $2\pi\times3.0$~kHz                                & $2\pi\times1.4$~kHz          \\
		$\tau$                            & $50~\mu$s                                          & $100~\mu$s                   \\
		\arrayrulecolor{lightgray}\hline\arrayrulecolor{black}
		$1/\sqmax$                        & 0.13                                               & 0.28                         \\
		Photon shot noise                 & 0.05                                               & 0.05                         \\
		%Imaging infidelity & 0.04 & 0.04\\
		Coupling variation                & 0.08                                               & 0.08                         \\
		Interaction strength noise        & 0.02                                               & 0.02                         \\
		Cavity photon loss                & 0.14                                               & 0.07                         \\
		%Free space scattering & 0.03 & 0.11\\
		Free space scattering             & 0.02                                               & 0.07                         \\ \arrayrulecolor{lightgray}\hline\arrayrulecolor{black}
		Measured contrast $C_\text{Rabi}$ & 0.89                                               & 0.89                         \\\hline
		Expected $\sq$                    & 0.44                                               & 0.56                         \\
		Observed $\sq$                    & $0.52\pm 0.07$                                     & $0.56\pm0.09$                \\
		\hline
	\end{tabular}
	\caption{Summary of noise sources contributing to the variances in Figs.~\ref{fig:2_global} and \ref{fig:two_cluster} of the main text, along with the relevant experimental parameters from Extended Data Table \ref{tab:figure_parameters}.  The method of calculating the expected variance from the individual noise contributions is described in Sec.~\ref{sec:noise_summary}.}
	\label{tab:squeezing_limits}
\end{table}

In principle, working at larger atom number increases the collective cooperativity, decreasing the relative effects of cavity dissipation. However, in addition to the noise sources in Table~\ref{tab:squeezing_limits}, different collective modes are sensitive to technical noise in the readout procedure, as discussed in Sec.~\ref{sec:imaging_calibration}. The example data presented in Table~\ref{tab:squeezing_limits} are measured in the $\up\down\up\down$ mode, which has negligible technical noise (see Extended Data Fig.~\ref{supp_fig:ShotNoise}).  For Figs.~\ref{fig:2_global} and \ref{fig:two_cluster} of the main text we squeeze up to two collective modes, and these modes can always be mapped to the two collective modes with minimal technical noise using local Larmor rotations. In this case we choose an atom number of $N = 1.5\times 10^4$, which is limited by the density of atoms in the trap, as any collisional spin exchange interactions are incoherent with the photon-mediated interactions, reducing the effective spin length. For the square graph state all 4 collective modes need to be squeezed, and the technical noise in our projection noise calibration becomes a relevant parameter, so we reduce the atom number in Fig.~\ref{fig:cluster_state} of the main text to $N = 8\times10^3$.

\subsection{Fundamental Scaling}
\label{sec:fundemental_scaling}
Fundamental limits to the degree of squeezing attainable by global cavity-mediated interactions among $N$ atoms are governed by the collective cooperativity $N\eta$~\cite{sorensen2002entangling}, where $\eta = 4g^2/(\kappa\Gamma)$ is the single-atom cooperativity.  In this section, we derive limits on squeezing multiple collective modes which demonstrate the scalability of our graph-state preparation protocol.  We focus first on the specific case of the spin-nematic squeezing employed in this work, and additionally comment on generalizations to other methods of cavity-mediated spin squeezing.

The fundamental limit on squeezing set by the cavity cooperativity arises from two dissipation processes: loss of photons from the cavity mode that mediates interactions; and scattering of photons into free space.  These loss processes are parameterized by the rates $\Gammac$ and $\Gamma_\text{sc}$ in Eq.~\eqref{eq:cavity_loss_params}.
The effect of the free-space scattering is proportional to the number of modes $M$ because scattering at any point during the protocol can reduce squeezing. Conversely, collective decay does not depend on $M$ because it acts only on the collective mode coupled to the cavity. To place free-space scattering and cavity loss on an even footing, we imagine dividing the squeezing of each collective mode into multiple segments interleaved with squeezing of the other collective modes, so that the scattering is interspersed with the coherent dynamics. Any scattering loss while addressing each of the $M$ modes should then be included as a component of Eq.~\eqref{eq:dsqdt_2ph}. The full equation of motion for the variance of each collective mode is thus
\begin{equation}
	\frac{d \sqmin}{dt} = -2\lambda \sqmin + \Gammac + 2M \Gamma_\text{sc} .
	\label{eq:full_loss}
\end{equation}
The squeezing is optimized by choosing the drive-cavity detuning that minimizes the total contribution to Eq.~\eqref{eq:full_loss} from scattering and cavity decay:
\begin{equation}
	\left(\frac{\delta_-}{\kappa}\right)^2 = \frac{1}{192} \frac{N\eta}{M}.
\end{equation}
This optimum detuning is set by the collective cooperativity per mode $N\eta/M$ and leads to an overall variance
\begin{equation}\label{eq:coop_limit}
	\sqmin = 8\left(\frac{N\eta}{3M}\right)^{-1/2}.
\end{equation}
While these derivations focus on variance $\sq$, the same scaling applies to the Wineland squeezing parameter $\xi^2$.

Equation~\eqref{eq:coop_limit} shows that our protocol for generating arbitrary $M$-node graph states by squeezing $M$ collective modes yields fixed squeezing at constant atom number per ensemble $(N/M)$, independent of the number of graph nodes. The protocol requires order $M$ squeezed modes and order $M$ sets of rotations.  However, at fixed atom number per ensemble, with increasing $M$ the increased collective interaction strength causes each mode to be squeezed faster, so that the total interaction time remains fixed.  The protocol can thus be scaled to larger arrays of ensembles, limited only by the spatial extent of the cavity mode and the fidelity of local addressing.

The limit on squeezing $\xi^2\propto 1/\sqrt{(N/M)\eta}$ set by the collective cooperativity per mode generalizes to a wide variety of methods of cavity spin squeezing, including approaches employing either photon-mediated interactions~\cite{sorensen2002entangling,leroux2010implementation} or quantum non-demolition measurements~\cite{hosten2016measurement}.  Improvements to both the numerical prefactor and the overall scaling with cooperativity are possible, however, by a suitable choice of atomic level scheme.  Notably, for squeezing on a cycling transition, the scaling for the single-mode case improves to $\xi^2\propto1/(N\eta)$~\cite{greve2022entanglement, bohnet2014reduced}.  Optimizing the scheme for the collective entangling operations may facilitate future work seeking the error-correction threshold of $-10\log\xi^2 = 20.5$ dB \cite{menicucci2014fault}, or generating discrete-variable graph states in arrays of single atoms.

\end{document}